\documentclass[10pt, twocolumn]{article}

\usepackage[a4paper, total={180mm, 250mm}, left=15mm, right=15mm, top=25mm, bottom=25mm]{geometry}
\usepackage[utf8]{inputenc}
\usepackage{authblk}
\usepackage{graphicx} 
\usepackage{float} 
\usepackage{booktabs}
\usepackage{longtable}
\usepackage{xcolor} 
\usepackage{hyperref}
\hypersetup{
    colorlinks=true,  
    linkcolor=blue,   
    urlcolor=blue,    
    citecolor=.,      
}
\usepackage{rotating}
\usepackage{array}
\usepackage{makecell}
\usepackage{placeins}

\newcolumntype{L}[1]{>{\raggedright\arraybackslash}p{#1}}

\usepackage[table]{xcolor}

\definecolor{rowgray}{RGB}{242,242,242} 

\usepackage[
    backend=biber,
    style=numeric,
    sorting=none,
    defernumbers=true,
    doi=false,      
    isbn=false,     
    url=false,      
    eprint=false    
]{biblatex}
\AtEveryBibitem{%
  \clearfield{urldate}
  \clearfield{note}
  \clearfield{issn}%
  \clearfield{isbn}%
  \clearfield{eprint}%
  \clearfield{publisher}%
  \clearfield{location}%
  \clearname{editor}
}
\renewbibmacro{in:}{}

\usepackage{tocloft}

\newfloat{extendedfigure}{htbp}{loe}

\floatname{extendedfigure}{Extended Data Figure}

\usepackage[labelfont=bf]{caption}
\usepackage{amsmath} 
\usepackage{amsfonts} 
\DeclareCaptionType{Video}[Video][video]

\newcommand{\iim}{\text{i}}  
\newcommand{\vect}[1]{\mathbf{#1}}  
\newcommand{\starr}{\star}  
\newcommand{\rvec}{\vect{r}}  
\newcommand{\nuvec}{\boldsymbol{\nu}}  

\DeclareMathOperator*{\argmin}{argmin}

\title{WaveOrder: A differentiable wave-optical framework for scalable biological microscopy with diverse modalities}

\author[1,*]{Talon~Chandler}
\author[1]{Ivan~E.~Ivanov}
\author[1]{Gabriel~Sturm}
\author[1]{Sheng~Xiao}
\author[1]{Xiang~Zhao}
\author[1]{Alexander~Hillsley}
\author[1]{Allyson~Quinn~Ryan}
\author[1,2]{Ziwen~Liu}
\author[1]{Sricharan~Reddy~Varra}
\author[1]{Ilan~Theodoro}
\author[1]{Eduardo~Hirata-Miyasaki}
\author[1]{Deepika~Sundarraman}
\author[1]{Amitabh~Verma}
\author[1]{Madhurya~Sekhar}
\author[1]{Chad~Liu}
\author[1]{Soorya~Pradeep}
\author[1]{See-Chi~Lee}
\author[3]{Shannon~N.~Rhoads}
\author[3]{Maria~Clara~Zanellati}
\author[3]{Sarah~Cohen}
\author[1]{Carolina~Arias}
\author[1]{Manuel~D.~Leonetti}
\author[1]{Adrian~Jacobo}
\author[1]{Keir~Balla}
\author[1]{Loïc~A.~Royer}
\author[1,*]{Shalin~B.~Mehta}

\affil[1]{Biohub, San Francisco, CA, USA}
\affil[2]{Center for Scalable Data Analytics and Artificial Intelligence, TU Dresden, Dresden, Germany}
\affil[3]{Cell Biology and Physiology, University of North Carolina, Chapel Hill, NC, USA}
\affil[*]{Correspondence: \texttt{\{talon.chandler,shalin.mehta\}@biohub.org}}

\begin{document}

\maketitle

\begin{abstract}
Correlative computational microscopy can accelerate imaging and modeling of cellular dynamics by relaxing trade-offs inherent to dynamic imaging. Existing computational microscopy frameworks are either specialized or overly generic, limiting use to fixed configurations or domain experts. We introduce \emph{WaveOrder}, a generalist \emph{wave}-optical framework for imaging the architectural \emph{order} of biomolecules. WaveOrder reconstructs diverse specimen properties from multi-channel acquisitions, with or without fluorescence. It provides a unified representation of linear optical properties and differentiable physics-based image formation models spanning widefield, confocal, light-sheet, and oblique label-free geometries. WaveOrder uses physics-informed ML to auto-tune model parameters and solve blind shift-variant restoration problems. This open-source, PyTorch-based framework enables scalable quantitative imaging across scales from organelles to adult zebrafish, and improves restoration of cellular structures in high-throughput experiments. We validate WaveOrder on diverse imaging applications, demonstrating its ability to recover biomolecular structure beyond the limits of existing approaches.
\end{abstract}




\begin{refsegment}
\section{Introduction}


Biological functions emerge from interactions among components that span length scales, including biomolecules, organelles, cells, tissues, and organs. Capturing the composition and organization of these components therefore requires correlative, high-throughput imaging across spatial, temporal, and molecular dimensions. However, conventional imaging systems are constrained by fundamental trade-offs between spatial resolution, temporal resolution, number of channels, field of view, and sample health, which limit their ability to probe biological processes at scale. Computational imaging methods, which can optically encode multiple physical and molecular properties and decode them computationally, offer a promising route to relaxing these trade-offs. To fully realize this potential, there is a need for a flexible computational framework that supports quantitative reconstruction of biomolecular distributions across diverse imaging configurations at high throughput. This paper introduces a computational imaging framework designed to meet this need.


A generalist computational imaging framework does more than relax tradeoffs imposed by dynamic imaging, it enables correlative microscopy by combining complementary contrast mechanisms through a common physical model. Correlative label-free and fluorescence imaging bridges dense structural information with sparse molecular labeling, linking physical properties of cellular compartments to molecular phenotypes~\cite{schlusler_correlative_2022} and supporting downstream tasks including virtual staining~\cite{guo_revealing_2020, gomez-de-mariscal_harnessing_2024, liu_robust_2025}. Dynamic correlative imaging extends these benefits to time-resolved studies of cellular responses across perturbations~\cite{ivanov_mantis_2024}. Similar strategies appear across modalities: structural and functional OCT (optical coherence tomography) combines scattering, elastography, spectroscopy, and multiphoton excitation to monitor cell dynamics and mechanics in engineered tissues~\cite{liang_imaging_2009}; multi-modal label-free metabolic imaging combines two-photon fluorescence, lifetime, and spectral readouts to quantify redox state and mitochondrial function in 3D brain models~\cite{zhang_multi-modal_2025}; and at finer scales, cryo-CLEM (correlative light and electron microscopy) combines molecular specificity from fluorescence with ultrastructure from EM to reveal mechanisms inaccessible to any single modality \cite{serwas_getting_2021, pierson_recent_2024}. These examples highlight a consistent need for unified, physics-based frameworks capable of modeling and inverting diverse contrast mechanisms so heterogeneous imaging data can be compared and interpreted quantitatively.

    
Many works describe microscopic image formation for individual contrast modes including fluorescence contrast~\cite{backer_extending_2014, chandler_spatio-angular_2019}, phase and absorption contrast~\cite{streibl_three-dimensional_1985, bao_quantitative_2016}, and polarization-resolved birefringence contrast~\cite{torok_imaging_2000, oldenbourg_point-spread_2000}. Several recent works have developed computational imaging methods for multi-channel measurements of specimen properties, but with models specific to label-free contrast~\cite{guo_revealing_2020, saba_polarization-sensitive_2021, shin_tomographic_2022, yeh_permittivity_2024, jang_super-resolution_2023} or fluorescence contrast~\cite{zhang_six-dimensional_2023, chandler_volumetric_2025}. The WaveOrder framework unifies and extends these models to \textit{multi-contrast} and \textit{multi-channel} imaging systems that include scalar and polarization-resolved imaging, with or without fluorescence labels\footnote{In this paper, \emph{multi-contrast} implies different light-matter interactions, e.g., label-free and fluorescence contrast, while \emph{multi-channel} implies channels acquired with a given contrast, e.g., multiple fluorophores or label-free phase and polarization microscopy.}.

The most general image formation models account for statistical fluctuations in the electric field of light, i.e., coherence, and how coherence is modulated due to propagation or interaction with matter. Modeling coherence requires bilinear functions (functions whose output depends on pairs of points in the imaging path) and propagation of second-order statistics~\cite{streibl_three-dimensional_1985, hopkins_concept_1951, barrett_foundations_2004, mehta_partially_2018, sheppard_partially_2018, wolf_introduction_2007}. We start with this approach before restricting our attention to imaging problems that can be modeled with linear functions. Specifically, we consider spatially incoherent fluorescence samples and label-free imaging systems with a spatially incoherent source. The optical transformation of the specimen's properties in many widely used imaging systems, such as widefield, confocal, light-sheet, quantitative phase imaging, and quantitative polarization imaging, are well approximated by linear image formation models.

Several existing frameworks provide some, but not all, of the features of WaveOrder (\textbf{Ext.\,Data\,Table\,\ref{exttab:method_comparison}}). Deconvolution libraries typically focus on fluorescence deconvolution~\cite{bruce_real-time_2013, sage_deconvolutionlab2_2017, li_incorporating_2022, wernersson_deconwolf_2024, alshaabi_fourier-based_2025, kohli_ring_2025}, limiting their value in multi-contrast correlative settings. Differentiable microscopy libraries~\cite{herath_differentiable_2022, deb_chromatix_2025, tachella_deepinverse_2025} flexibly model a wide variety of imaging systems to enable new designs and reconstructions, but they typically require a domain expert to match them to a specific application. WaveOrder prioritizes linear reconstruction models for the most widely used computational microscopy contrast methods, facilitating broad applications.  

Although deconvolution methods are widely used in microscopy, in practice it is often unclear what point-spread or transfer function to use, making reconstruction a blind deconvolution problem. Blind-deconvolution is even more relevant for high-throughput or large-scale imaging experiments where contrast is shift-variant, as illumination, aberrations, and depth-dependent distortions vary across large fields of view. Classical blind deconvolution has been extensively applied to natural images \cite{levin_understanding_2009}, and learning-based approaches are increasingly applied to microscopy data \cite{gilton_model_2021, kang_coordinate-based_2024, alshaabi_fourier-based_2025, volpe_roadmap_2025}. WaveOrder builds on these developments by contributing a physics-guided differentiable computational graph that reduces the dimension of the blind search space and enables auto-tuning of imaging parameters directly from data. 

We also find that many reconstruction implementations are not used broadly for biological research because they are not reproducible or easy to access. WaveOrder provides an open-source implementation of linear computational imaging methods that unifies widely used models and reconstruction algorithms within a common framework. WaveOrder leverages PyTorch for cross-platform, high-performance analysis of large datasets and integration with learned computer vision models.

In our view, this paper makes the following contributions
\begin{itemize}\itemsep0em
    \item a unified and elegant mathematical representation of material properties, illumination, scattering, and detection across a broad range of microscopy imaging systems as a computational graph;
    \item a demonstration of reconstructions of biological samples across length scales from organelles to organisms, including auto-tuning of reconstruction parameters with shift-variant data-dependent image quality loss;
    \item a simulation and reconstruction framework that can be applied to linear microscopy contrast modes, including fluorescence, phase, absorption, birefringence, and diattenuation; and 
    \item a differentiable PyTorch library and scalable pipeline that democratizes computational imaging for both developers and users of diverse imaging systems.  
\end{itemize}

Next, we report how WaveOrder models diverse contrast mechanisms, reconstructs volumetric datasets from simulation and experiments, and uses physics-informed machine learning to auto-tune model parameters. 

\section{Results}\label{supp:results}
\subsection{Computational-graph formulation enables simulation and inversion of microscopy data}

\begin{figure*}
    \centering
    \includegraphics[width=180mm]{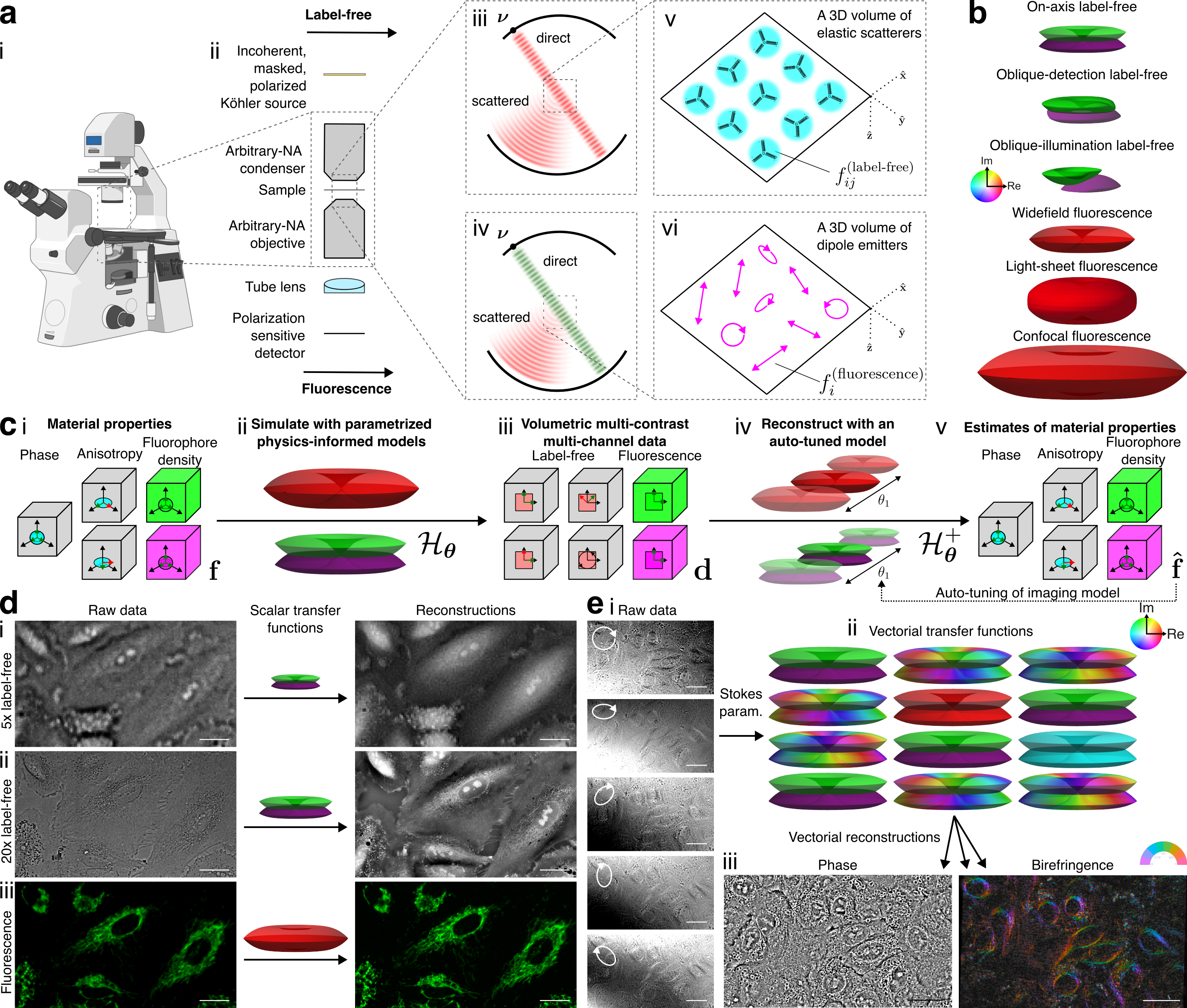}
    \caption{\textbf{WaveOrder models and reconstructs multi-contrast volumetric microscopy data.} \textbf{(a)} The framework models imaging with \textbf{(i)} microscopes equipped with \textbf{(ii)} an incoherent source with a programmable amplitude mask, tunable spectral/polarization filters, and a polarization-sensitive detector. Label-free contrast arises from \textbf{(iii)} direct-scatter interference, and fluorescence contrast from \textbf{(iv)} filtered scatter-scatter interference. Coherent scatterers are represented as \textbf{(v)} a volume of scattering potential tensors, and fluorescent emitters as \textbf{(vi)} a volume of dipole emission moments. \textbf{(b)} Parametrized transfer functions generalize across microscope geometries and contrast types. Transfer function colors indicate the relative phase of complex values, see inset color wheel. \textbf{(c)} WaveOrder consists of \textbf{(i)} representations of material properties $\mathbf{f}$ operated on by \textbf{(ii)} parametrized, physics-informed models $\mathcal{H}_{\boldsymbol{\theta}}$ to simulate \textbf{(iii)} multi-contrast multi-channel volumetric data $\mathbf{d}$. WaveOrder applies \textbf{(iv)} pseudo-inverse operators $\mathcal{H}_{\boldsymbol{\theta}}^{+}$ to \textbf{(v)} estimate specimen properties. Transfer function parameters  $\boldsymbol{\theta}$ and the specimen properties $\mathbf{\hat{f}}$ are refined iteratively. \textbf{(d)} WaveOrder restores scalar datasets (e.g.\,\textbf{(i)} label-free at 5$\times$, \textbf{(ii)} 20$\times$, and \textbf{(iii)} confocal fluorescence) and \textbf{(e)} vectorial datasets (e.g.\,\textbf{(i)} polarization-resolved acquisitions use \textbf{(ii)} a bank of transfer functions to recover \textbf{(iii)} phase and birefringence). \textbf{Scale bars:} 25~\textmu m.}
    \label{fig:overview}
\end{figure*}

We first illustrate how WaveOrder models image formation and reconstruction in standard microscopes (\textbf{Fig.\,\ref{fig:overview}a--c}). Imaging is represented as a computational graph that links the sample, illumination, scattering, and detection with composable operators, enabling direct simulation and inversion (\textbf{Methods\,\ref{sec:framework}}). Within this formulation, WaveOrder reproduces label-free contrast by simulating interference between direct and scattered fields (\textbf{Fig.\,\ref{fig:overview}a, iii}) and fluorescence contrast by simulating inelastic, spectrally shifted emission (\textbf{Fig.\,\ref{fig:overview}a, iv}). 

WaveOrder represents specimen properties and detected light as physically interpretable vectors (\textbf{Ext.\,Data\,Fig.\,\ref{extfig:basis-functions},\,Methods\,\ref{sec:objects_and_data}}). For label-free specimens, WaveOrder uses scattering potential tensors expanded onto spherical harmonic tensors, with familiar expansion coefficients: phase, absorption, birefringence, and diattenuation. For fluorescent specimens, WaveOrder uses dipole moment vectors and their second moments expanded onto the spherical harmonics. For polarization-resolved detected light, WaveOrder uses the Stokes parameters (\textbf{Methods\,\ref{sec:interpretable_basis},\,Supp.\,3--4}).

WaveOrder models the relationships between specimen properties and detected light as linear operators called transfer functions (\textbf{Fig.\,\ref{fig:overview}b,\,Methods\,\ref{sec:linear_operators}}). All transfer functions in WaveOrder are constructed from three core submodels: a scattering model, an illumination pupil, and a detection pupil (\textbf{Methods\,\ref{sec:tf_summary},\,Supp.\,5}). WaveOrder uses a single scattering model, the Green's tensor spectrum, which describes how dipole emitters radiate polarized fields (\textbf{Ext.\,Data\,Fig.\,\ref{extfig:transfer}a--b,\,Supp.\,5.2}). When combined with an illumination and detection pupil, these submodels can generate scalar and vector transfer functions in both label-free and fluorescence contrast (\textbf{Table\,\ref{tab:tf},\,Ext.\,Data\,Fig.\,\ref{extfig:transfer}c--g,\,Methods\,\ref{sec:unified}}). This modular construction forms the core of WaveOrder's computational graph, allowing us to simulate (\textbf{Fig.\,\ref{fig:overview}b,\,i--iii}) and invert (\textbf{Fig.\,\ref{fig:overview}b,\,iii--v;\,Methods\,\ref{sec:reconstructing_high}}) diverse microscope geometries by varying a small number of pupil parameters.  

Parametrized transfer functions generalize these models across geometries and contrast mechanisms (\textbf{Fig.\,\ref{fig:overview}b}), providing a unified representation of multi-contrast, multi-channel imaging systems that can be inverted to recover phase, absorption, birefringence, fluorescence density, and fluorescence orientation (\textbf{Fig.\,\ref{fig:overview}d--e,\,Videos\,3--4}).

\subsection{Physics-guided ML auto-tunes multi-contrast blind deconvolution}
\begin{figure*}
  \centering
  \includegraphics[width=\textwidth]{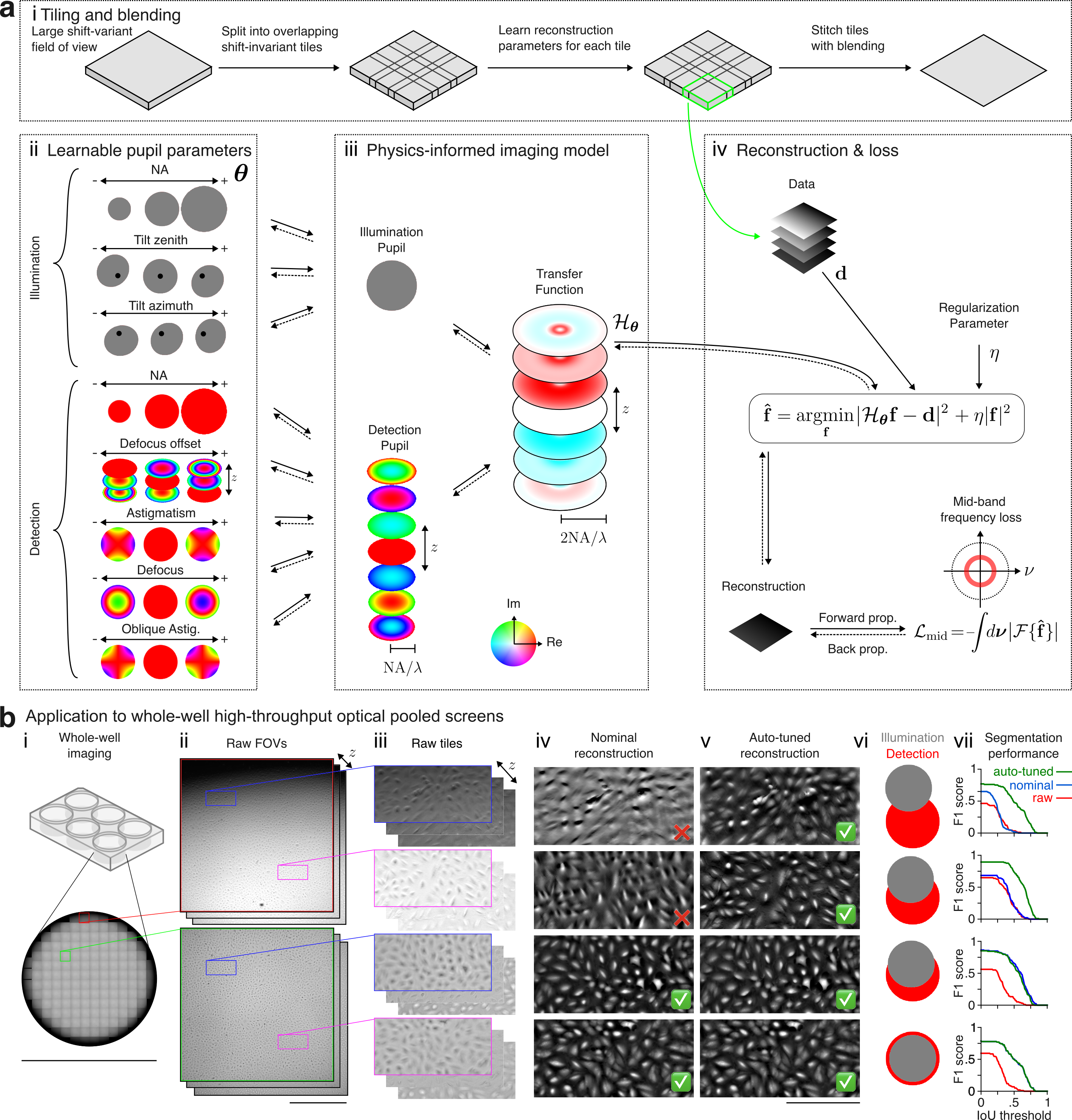}
  \caption{\textbf{Tiling and auto-tuning enable shift-variant blind deconvolutions.} Large fields of view often exhibit shift-variant contrast because of variations in the transfer functions, making reconstruction a blind deconvolution problem. We address this by \textbf{(a)(i)} dividing the field of view into overlapping, approximately shift-invariant tiles. We reconstruct each tile using a physics-guided imaging model \textbf{(ii)} parameterized by a vector $\boldsymbol{\theta}$ describing optical misalignments and aberrations (illustrated for eight common misalignments). From these parameters we compute \textbf{(iii)} transfer functions $\mathcal{H}_{\boldsymbol{\theta}}$ (illustrated for phase-from-defocus with its illumination and detection pupils). We use the model as input to a \textbf{(iv)} Tikhonov-regularized least-squares problem to estimate $\mathbf{\hat{f}}$ in a single step. We compute a scalar loss on the reconstruction (illustrated for a mid-band spatial frequency loss) and update $\boldsymbol{\theta}$ by backpropagation. \textbf{(b)} We demonstrate on optical-pooled screen data from \textbf{(i)} a full 35 mm well. \textbf{(ii--iii)} Edge tiles show oblique-illumination contrast while central tiles show on-axis contrast. \textbf{(iv)} Nominal reconstructions fail on edge tiles, while  \textbf{(v)} auto-tuned reconstructions recover consistent contrast and \textbf{(vi)} simultaneously estimate the underlying illumination tilt. \textbf{(vii)} CellPose segmentation performed best on the auto-tuned reconstruction. \textbf{Scale bars: (b, i)} 35~mm; \textbf{(b, ii)} 1~mm; \textbf{(b, iv--v)} 250~\textmu m. Pupil and transfer function colors indicate the relative size and phase of complex values, see color wheel in \textbf{(a, iii)}.}
  \label{fig:tiles}
\end{figure*}

Large fields of view often exhibit uncontrolled, shift-variant contrast, turning reconstruction into many blind deconvolution problems, where the object and the point-response function need to be simultaneously estimated. We address this by dividing the field of view into overlapping, approximately shift-invariant tiles (\textbf{Fig.\,\ref{fig:tiles}a,\,i}) and associating each tile with a parameter vector $\boldsymbol{\theta}$ that captures illumination and detection misalignment and aberration (\textbf{Fig.\,\ref{fig:tiles}a,\,ii}). From $\boldsymbol{\theta}$ we compute tile-specific transfer functions $\mathcal{H}_{\boldsymbol{\theta}}$ (\textbf{Fig.\,\ref{fig:tiles}a,\,iii}) and reconstruct the object with a Tikhonov-regularized pseudo-inverse (\textbf{Methods\,\ref{sec:recon_low}}). Finally, we update $\boldsymbol{\theta}$ by backpropagating a scalar loss (e.g., mid-band spatial frequency power spectrum to encourage sharp features while controlling noise), updating our parameter and object estimates iteratively (\textbf{Fig.\,\ref{fig:tiles}a,\,iv;\,Ext.\,Data\,Fig.\,\ref{extfig:optimization-arch};\,Videos\,1--2}).

On optical pooled screen data spanning a full 35~mm well (\textbf{Fig.\,\ref{fig:tiles}b,\,i}), central tiles exhibit on-axis contrast while edge tiles show oblique illumination due to the meniscus effect at the edge of the well (\textbf{Fig.\,\ref{fig:tiles}b,\,ii--iii}). Nominal (untuned) reconstructions fail at the periphery, producing blurred or inverted contrast (\textbf{Fig.\,\ref{fig:tiles}b,\,iv}). Auto-tuning restores consistent contrast across tiles (\textbf{Fig.\,\ref{fig:tiles}b,\,v}) and simultaneously estimates the underlying illumination tilt (\textbf{Fig.\,\ref{fig:tiles}b,\,vi}), $\sim$$12^{\circ}$ tilt for the most peripheral tiles (top row,\,\textbf{Fig.\,\ref{fig:tiles}b}). We quantified these improvements by benchmarking CellPose segmentations against manual annotations for raw, nominal, and auto-tuned reconstructions. Mean F1 scores at an IoU threshold of 0.5 increased from 0.10 (raw) to 0.35 (nominal) and 0.58 (auto-tuned), with the largest gains observed on peripheral tiles (\textbf{Fig.\,\ref{fig:tiles}b,\,vii;\, Ext.\,Data\,Fig.\,\ref{extfig:ops-segment},\,Methods\,\ref{sec:ops}}). 

\subsection{Scalar reconstructions enable biological insight across scales}

WaveOrder restores quantitative phase and fluorescence volumes for datasets spanning scales from organelles to organisms (\textbf{Figs.\,\ref{fig:quantitative_reconstruction}--\ref{fig:zfish_reconstruction}}). By modeling image formation and flexibly auto-tuning the transfer function for individual datasets, WaveOrder improves signal-to-noise ratio, removes defocus ambiguity, and enables consistent physical interpretations. Together, these capabilities allow WaveOrder to extract biologically meaningful structure across imaging modalities, microscope geometries, and length scales. 

\subsubsection{Correlative imaging of organelles, cells, and organs}
\begin{figure*}
    \centering
    \includegraphics[width=\textwidth]{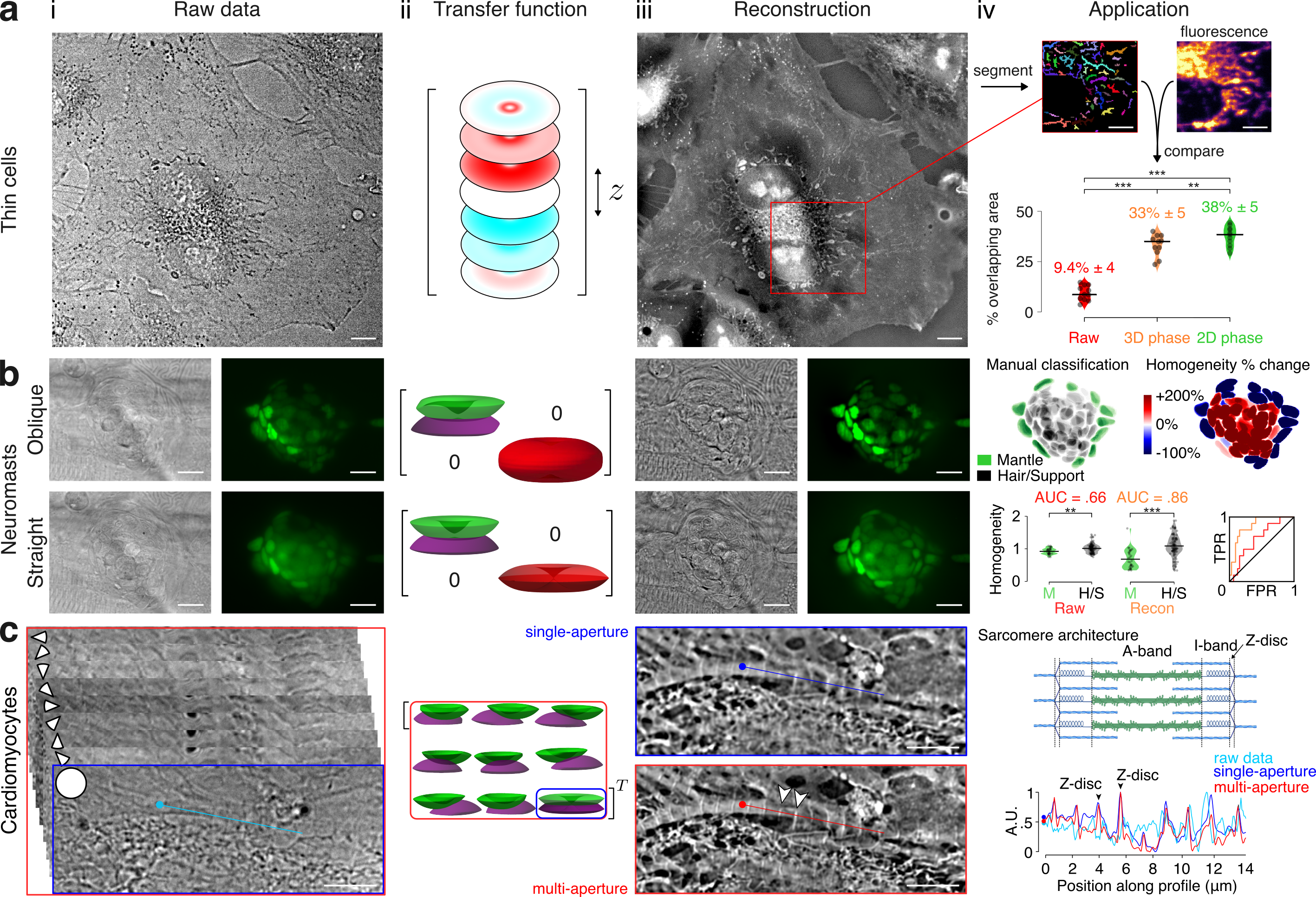}
    \caption{\textbf{Flexible scalar reconstructions enable diverse biological applications.} \textbf{(a)(i)} Thin label-free cells imaged in transmission suffer from low contrast, in-focus disappearance, and ambiguity in interpreting intensity values. Acquiring a through-focus stack together with \textbf{(ii--iii)} WaveOrder's auto-tuned reconstructions enables sharp single-plane reconstructions that summarize out-of-focus information, improve SNR, and remove the sign ambiguity of raw data. \textbf{(iv)} Comparing thresholded Frangi-filtered mitochondria segmentations from label-free and simultaneously acquired fluorescent mitochondrial imaging demonstrates the improved segmentation performance of a single-plane phase reconstruction. \textbf{(b)(i)} Zebrafish neuromasts imaged using (top) light-sheet illumination and oblique-detection configuration, typical of oblique-plane microscope geometries, yields contrast similar to a more typical (bottom) on-axis label-free and widefield fluorescence configuration. \textbf{(ii--iii)} WaveOrder enables reconstructions of all four contrast types, producing sharpened, higher-SNR images across modalities. \textbf{(iv)} Manual classification of mantle versus hair/support cells (top left) is similar to the fluorescence homogeneity \% change after reconstruction (top right). Violin plots (bottom left) show improved separation between groups after reconstruction, with ROC curves (bottom right) confirming the improved classification. \textbf{(c)(i)} Cardiomyocytes imaged under nine oblique illumination geometries do not clearly reveal sarcomere architecture in the raw data alone. \textbf{(ii--iii)} WaveOrder's reconstruction routines enable visualization of z-discs from a single volume (blue outlines) and show further contrast enhancement when all nine illuminations are combined into a single phase estimate (red outlines). Transfer functions are in a single $9\times 1$ column vector, arranged into a $3\times 3$ grid. \textbf{(iv)} Sarcomere architecture (top, created with BioRender) and profiles (bottom) through the raw data (light blue), single-aperture reconstruction (dark blue), and multi-aperture reconstruction (red) demonstrate the improved contrast. \textbf{Scale bars: (a--b)} 10 \textmu m; \textbf{(c) }5~\textmu m.}
    \label{fig:quantitative_reconstruction}
\end{figure*}

In thin, adherent A549 cells, raw label-free images exhibit low contrast and in-focus disappearance due to destructive interference of direct and scattered fields (\textbf{Fig.\,\ref{fig:quantitative_reconstruction}a,\,i;\,Ext.\,Data\,Fig.\,\ref{extfig:light-paths}a}). Acquiring a through-focus stack and applying WaveOrder's auto-tuned scalar reconstruction routines produces a sharp single-plane image that summarizes out-of-focus information and produces a whole-cell phenotype (\textbf{Fig.\,\ref{fig:quantitative_reconstruction}a,\,ii--iii}), improving SNR and restoring a direct monotonic mapping between intensity and density. Comparing thresholded Frangi-filtered segmentations from raw fluorescence data to segmentations from auto-tuned label-free reconstructions (\textbf{Fig.\,\ref{fig:quantitative_reconstruction}a,\,iv};\,\textbf{Ext.\,Data\,Fig.\,\ref{extfig:organelle-segmentation}a;\,Methods\,\ref{sec:organelle}}) showed improved pixel-wise recall of mitochondrial segmentations (38\% $\pm$ 5) compared to raw data (9.4\% $\pm$ 4), demonstrating the phenotypic and biological value of physics-informed reconstructions (mean $\pm$ s.d.). Endosome segmentations (\textbf{Ext.\,Data\,Fig.\,\ref{extfig:organelle-segmentation}b}) show similar improvements.

Next, we assessed WaveOrder's performance on the zebrafish neuromast (\textbf{Fig.\,\ref{fig:quantitative_reconstruction}b,\,Methods\,\ref{sec:neuromast}}), a multi-cellular sensory structure acquired under oblique and straight detection geometries (\textbf{Ext.\,Data\,Fig.\,\ref{extfig:light-paths}a--b}) \cite{ma_signaling_2009, tang_wntbeta-catenin_2019}. Reconstructions recover high-SNR label-free and fluorescence images under both imaging configurations (\textbf{Fig.\,\ref{fig:quantitative_reconstruction}b,\,ii--iii}) demonstrating that the same transfer-function formalism generalizes across geometries. Using fluorescence contrast, we evaluated whether WaveOrder's reconstructions improved our ability to classify mantle cells versus hair/support cells using a single intensity-derived texture metric, homogeneity (\textbf{Fig.\,\ref{fig:quantitative_reconstruction}b,\,iv;\,Ext.\,Data\,Fig.\,\ref{extfig:neuromast-segmentation}}). Across three neuromasts, cell-level homogeneity values showed greater separation between classes after reconstruction. ROC curves indicated poor classification performance on raw data (AUC = 0.66), improving substantially after reconstruction (AUC = 0.86; \textbf{Fig.\,\ref{fig:quantitative_reconstruction}b,\,iv}). Traditionally, developmental cell types are distinguished by gene expression differences, but shape and texture features have become valuable discriminators recently \cite{chacon-martinez_signaling_2018, alves-afonso_tissue_2021}. While enhancer traps remain a powerful tool for mantle cell annotation \cite{viader-llargues_live_2018, brooks_pharmacological_2024}, accurate classification based on whole-organism fluorescence signal texture alone enables broader inference across cell types.

In cardiomyocytes imaged under nine oblique illumination angles (\textbf{Ext.\,Data\,Fig.\,\ref{extfig:light-paths}c,\,Methods\,\ref{sec:cardio}}), raw images weakly reveal sarcomere structures (\textbf{Fig.\,\ref{fig:quantitative_reconstruction}c,\,i}). Single-aperture reconstructions recover z-disc periodicity, and multi-aperture fusion further enhances contrast and sharpness (\textbf{Fig.\,\ref{fig:quantitative_reconstruction}c,\,ii--iii}). While increased frame averaging contributes to improved signal-to-noise, the enhanced periodic modulation is consistent with improved phase recovery from multi-aperture fusion. Line profiles confirm enhanced modulation at the expected sarcomere spacing (\textbf{Fig.\,\ref{fig:quantitative_reconstruction}c,\,iv}). Resolving sarcomere banding in label-free images of cardiomyocytes is valuable for assessing muscle health and structure, and WaveOrder achieves this from previously difficult-to-interpret raw intensity images.

We further applied WaveOrder's scalar reconstruction framework to joint label-free and multispectral fluorescence datasets from human iPSCs, recovering high-contrast phase maps that correlate with spectrally unmixed fluorescence channels across differentiation stages (\textbf{Ext.\,Data\,Fig.\,\ref{extfig:multispectral-ipsc}}).

\subsubsection{Correlative imaging of tissue and cell density across a zebrafish embryo}
\begin{figure*}
    \centering
    \includegraphics[width=\textwidth]{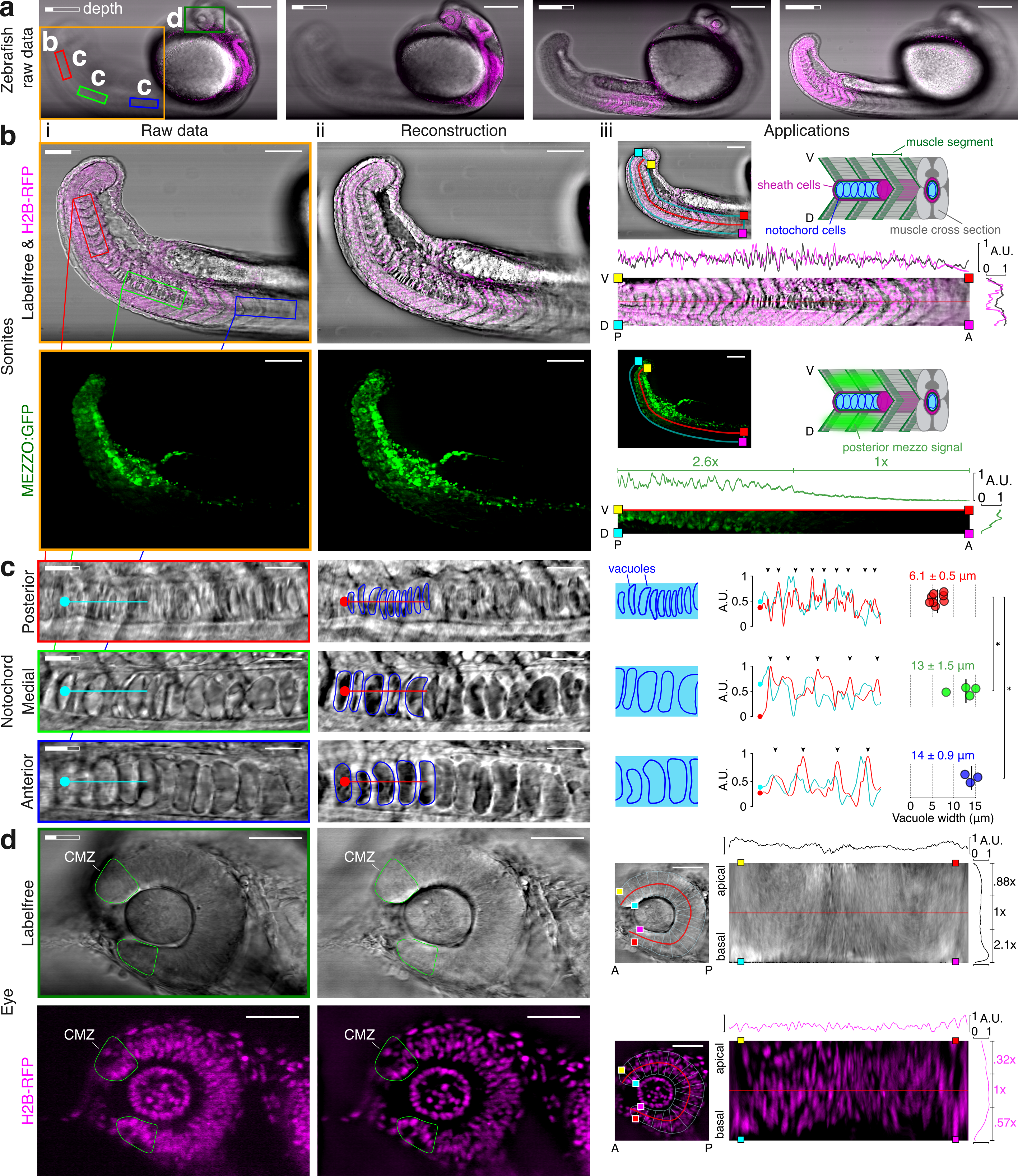}
    \caption{\textbf{Scalar reconstructions enable zebrafish phenotyping and characterization.} A living $\sim$20~hpf zebrafish embryo imaged with simultaneous light-sheet fluorescence and label-free contrast. 
    \textbf{(a)} Representative raw slices at different depths with ROIs indicated for panels \textbf{(b--d)}. 
    \textbf{(b)(i)} Tail, notochord, somites, and cellular structures visible in label-free and fluorescent channels; 
    \textbf{(ii)} reconstruction improves contrast in both modalities; 
    \textbf{(iii)} anatomical annotations enable tail straightening and quantification of Mezzo:GFP.
    \textbf{(c)(i)} Raw label-free data show ambiguous vacuole contrast; 
    \textbf{(ii)} reconstruction resolves vacuoles as low-density compartments; 
    \textbf{(iii)} single-profile measurements reveal a posterior-anterior increase in vacuole size.
    \textbf{(d)(i--ii)} Reconstruction highlights the retinal boundary and CMZ patterning in complementary channels; 
    \textbf{(iii)} annotations support computational unwrapping to assess basal-apical density differences. 
    \textbf{Scale bars:} \textbf{(a)} 100~\textmu m; \textbf{(b)} 50~\textmu m; \textbf{(c)} 25~\textmu m; \textbf{(d)} 50~\textmu m. 
    \textbf{Abbreviations:} \textbf{V}entral, \textbf{D}orsal, \textbf{P}osterior, \textbf{A}nterior.}
    \label{fig:zfish_reconstruction}
\end{figure*}

We extended the scalar reconstruction approach to a living $\sim$20~hpf (hours post fertilization) zebrafish embryo imaged with simultaneous light-sheet fluorescence and label-free contrast (\textbf{Fig.\,\ref{fig:zfish_reconstruction}a,\,Methods\,\ref{sec:zfishembryo}}), where raw sections show the tail, notochord, somites, and various cellular structures (\textbf{Fig.\,\ref{fig:zfish_reconstruction}b--d,\,i}). WaveOrder reconstructions improve contrast in both modalities (\textbf{Fig.\,\ref{fig:zfish_reconstruction}b--d,\,ii}), enabling anatomically guided unwrapping of the somites, notochord, and retina (\textbf{Fig.\,\ref{fig:zfish_reconstruction}b--d,\,iii}).

Turning our focus to the somites (\textbf{Fig.\,\ref{fig:zfish_reconstruction}b}), we imaged label-free + H2B-RFP (top) together with \textit{Tg(Mezzo:EGFP)} (bottom). In the raw data, Mezzo expression appears weak and diffuse (\textbf{Fig.\,\ref{fig:zfish_reconstruction}b,\,i}), but reconstruction substantially sharpens and localizes the signal (\textbf{Fig.\,\ref{fig:zfish_reconstruction}b,\,ii}). The improved label-free contrast simultaneously resolves somite and notochord boundaries, providing reliable anatomical landmarks for downstream analysis (\textbf{Fig.\,\ref{fig:zfish_reconstruction}b,\,iii}).

\textit{Tg(Mezzo:EGFP)} is a pan-mesendodermal reporter line in zebrafish. Mezzo is a paired-like homeobox protein and immediate target of Nodal signaling. The transgene shows cytoplasmic GFP expression in mesendoderm precursors from early gastrulation ($\sim$4 hpf) through late segmentation stages. Mezzo expression is directly correlated with somitogenesis; even in the late stage it remains expressed in the posterior-most presomitic mesoderm and tailbud, as long as somitogenesis continues posteriorly. It is widely used for live imaging of germ layer formation, cell migration dynamics during gastrulation, and studying endoderm specification \cite{poulain_mezzo_2002, lange_multimodal_2024}. 

With the reconstructed anatomical boundaries, we computationally straighten the tail and measure Mezzo:GFP intensity along the posterior-anterior axis (\textbf{Fig.\,\ref{fig:zfish_reconstruction}b,\,iii}). This analysis reveals a 2.6-fold ($\pm$ $3\times 10^{-4}$, s.e.) increase in posterior over anterior expression, consistent with the known persistence of Mezzo in the presomitic mesoderm and tailbud.

Building on the posterior-anterior trends observed in the Mezzo signal, we next analyzed notochord structure in the same region (\textbf{Fig.\,\ref{fig:zfish_reconstruction}c}). In raw label-free data, individual vacuoles show ambiguous sign and poorly defined boundaries (\textbf{Fig.\,\ref{fig:zfish_reconstruction}c,\,i}). Reconstruction resolves these vacuoles consistently as lower density compartments (\textbf{Fig.\,\ref{fig:zfish_reconstruction}c,\,ii}), enabling reliable single-profile measurements along the notochord (\textbf{Fig.\,\ref{fig:zfish_reconstruction}c,\,iii}). These measurements reveal a posterior-anterior increase in vacuole width, with mean sizes 6.1 \textmu m (posterior), 13 \textmu m (medial), and 14 \textmu m (anterior). 

The notochord is a defining feature of chordates that forms the primary embryonic axis. In zebrafish, specification begins during gastrulation ($\sim$6 hpf) and depends on transcription factors including \textit{floating head} (\textit{flh}/\textit{not}) and \textit{no tail} (\textit{ntl}), both essential for notochord precursor specification and differentiation. The notochord undergoes dramatic morphogenesis, forming a rod-like structure through mediolateral cell intercalation and vacuolation. By 24 hpf, the mature notochord consists of a single row of large vacuolated cells surrounded by extracellular matrix, providing structural support and signaling cues for neural tube and somite patterning. \cite{talbot_homeobox_1995, odenthal_mutations_1996}. The vacuole-size gradient we observe is consistent with this developmental progression and complements the posterior persistence of Mezzo expression, offering an independent label-free readout of maturation state.

Finally, we examined reconstruction-enabled improvements in other tissues, focusing on the developing eye (\textbf{Fig.\,\ref{fig:zfish_reconstruction}d}). The reconstructed fluorescence and label-free channels together reveal early organization of the ciliary marginal zone (CMZ). The fluorescence signal highlights a distinct shift in cell patterning at the retinal periphery, while the label-free channel provides a sharp boundary for the outer retinal surface. Combined, these modalities enable manual delineation of the CMZ and visualization of its internal organization (\textbf{Fig.\,\ref{fig:zfish_reconstruction}d,\,i--ii}). After dividing the retina into three equal regions along the apical-basal axis, quantification of the reconstructed phase shows a 2.1-fold ($\pm$ 0.1) higher density in the basal third of the retina compared to the middle third, particularly on the anterior side, indicating early structural heterogeneity within the forming CMZ (\textbf{Fig.\,\ref{fig:zfish_reconstruction}d,\,iii}).

The CMZ is a specialized germinal region at the peripheral rim of the zebrafish retina that contains retinal stem cells. The CMZ becomes morphologically distinct around 60 hpf ($\sim$2.5~days post fertilization (dpf)) and is functionally established by 72~hpf (3~dpf), after embryonic retinal differentiation is largely complete \cite{centanin_fate_2011}. Our imaging results show that cell organization in the CMZ is already distinct from other retinal areas as early as 1~dpf, and the observed basal-apical density differences are consistent with early compartmentalization of proliferative retinal stem cell niches. 

Together, these results demonstrate that scalar WaveOrder reconstructions enable improved phase and fluorescence phenotyping across cellular and organismal scales while preserving quantitative relationships between optical contrast and biological structure. 

\subsection{Vector multi-contrast computational microscopy from organelles to organisms}
\begin{figure*}
    \centering
    \includegraphics[width=\textwidth]{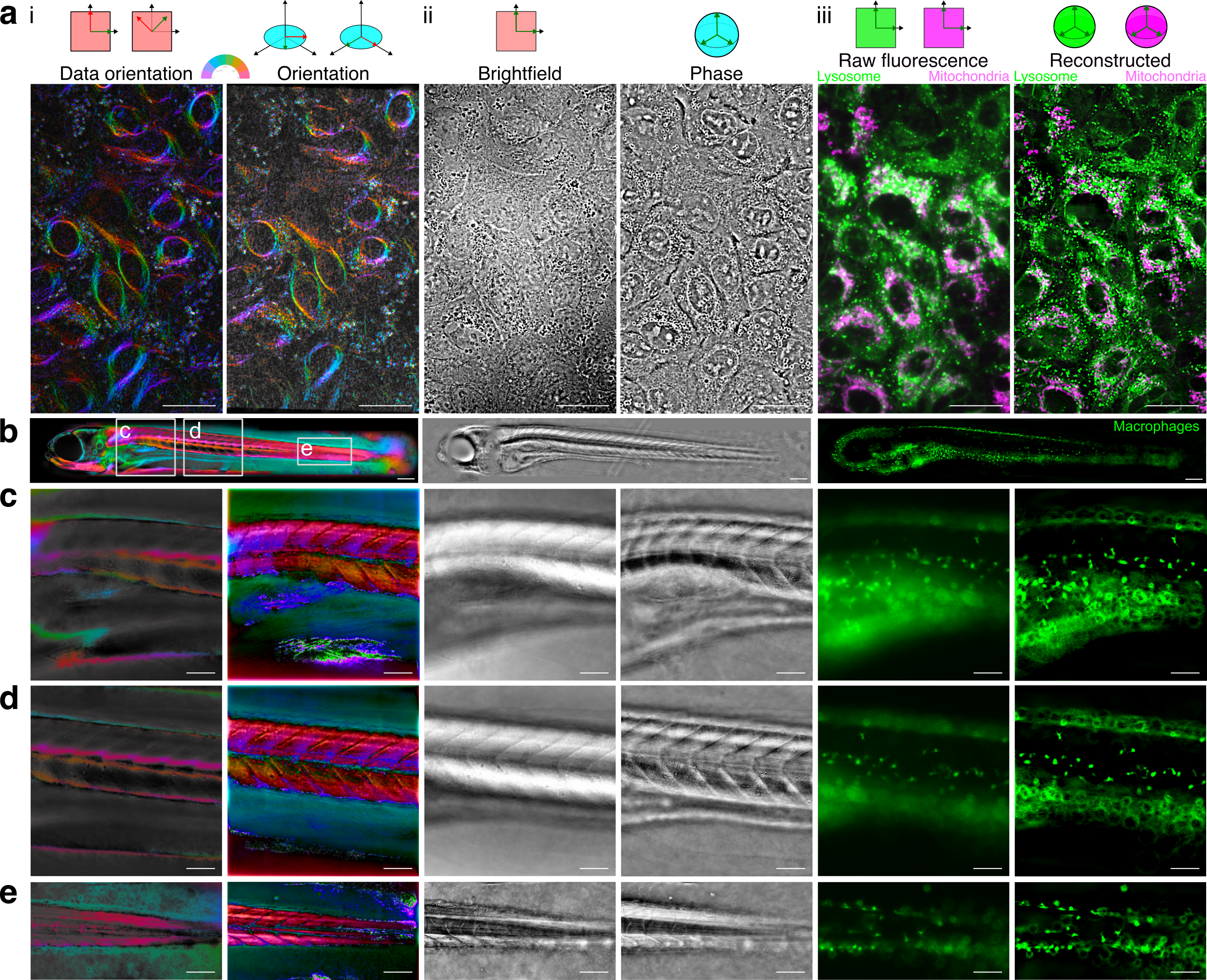}
    \caption{\textbf{Multi-contrast vectorial reconstructions improve visualization and interpretation across length scales} in \textbf{(a)} A549 cells, \textbf{(b)} zebrafish stitched from seven fields of view (edges feathered to reduce background and tile edges), and \textbf{(c--e)} zebrafish regions of interest. \textbf{(i)} Orientation data (left) and reconstructions (right). \textbf{(ii)} Brightfield data (left) and phase reconstructions (right). \textbf{(iii)} Fluorescence data (left) and reconstructions (right). See also \textbf{Videos 6 and 7}. \textbf{Scale bars: (a)} 25 \textmu m; \textbf{(b)} 200 \textmu m; \textbf{(c--e)} 100 \textmu m.} \label{fig:length_scales}
\end{figure*}

We applied WaveOrder reconstructions to multi-contrast, multi-channel data acquired from samples across length scales. \textbf{Fig.\,\ref{fig:length_scales}} shows alternating columns of data and reconstructions for transverse birefringence, phase, and fluorescence density. In data acquired from A549 cells (\textbf{Fig.\,\ref{fig:length_scales}a,\,Video\,6}) we observe improved sectioning, denoising, and contrast in phase and reconstructed fluorescence properties compared to their raw-data counterparts (\textbf{Fig.\,\ref{fig:length_scales}a,\,ii--iii}). In the orientation channel (\textbf{Fig.\,\ref{fig:length_scales}a,\,i}) we observe marginal improvements in contrast, but generally poor performance with reduced SNR and suppression of features that are apparent in the raw data. We attribute some of the performance drop to our imperfect noise model---we reconstruct from non-Gaussian Stokes parameters which is at odds with our Tikhonov least-squares reconstruction algorithm. Additionally, we have not explored the interaction between our Stokes-based background correction and our wave-optical reconstructions, another likely area for improvement.

We acquired multi-contrast data from an entire living zebrafish (\textbf{Fig.\,\ref{fig:length_scales}b}), then reconstructed fluorescence density, phase, and birefringence from specific regions of interest (\textbf{Fig.\,\ref{fig:length_scales}c--e,\,Video\,7}). We observe improved sectioning, denoising, and contrast in all three reconstructions. For example, in the label-free channel (\textbf{Fig.\,\ref{fig:length_scales}c--e,\,i}) we see improved contrast between the gut and muscles, and in the fluorescence channels (\textbf{Fig.\,\ref{fig:length_scales}c--e,\,iii}) we see improved contrast and resolution of immune cells. Improved contrast in the zebrafish gut (bottom of \textbf{Fig.\,\ref{fig:length_scales}c}) is particularly valuable for tracking immune-cell dynamics encoded in fluorescence reconstructions.

We validated and further tested WaveOrder's polarization-resolved multi-channel reconstructions on data acquired from a laser-etched anisotropy phantom with transverse radially anisotropic bubbles arranged in a spoke pattern (\textbf{Ext.\,Data\,Fig.\,\ref{extfig:vector_quantitative_reconstruction}a,\,i}). We measured four volumetric Stokes datasets (\textbf{Ext.\,Data\,Fig.\,\ref{extfig:vector_quantitative_reconstruction}a,\,ii}) and applied WaveOrder's reconstruction algorithm to estimate three label-free material properties that correspond to phase and transverse birefringence (\textbf{Ext.\,Data\,Fig.\,\ref{extfig:vector_quantitative_reconstruction}a,\,iii;\,Video\,5}).

In parallel, we simulated the anisotropic phantom (\textbf{Ext.\,Data\,Fig.\,\ref{extfig:vector_quantitative_reconstruction}b,\,i}) and the image formation process (\textbf{Ext.\,Data\,Fig.\,\ref{extfig:vector_quantitative_reconstruction}b,\,ii}), then we applied an identical reconstruction algorithm to estimate material properties (\textbf{Ext.\,Data\,Fig.\,\ref{extfig:vector_quantitative_reconstruction}b,\,iii}). \textbf{Ext.\,Data\,Figs.\,\ref{extfig:vector_quantitative_reconstruction}a,\,ii-iii} and \textbf{b,\,ii--iii} can be directly compared to indicate the quality of our models, where differences can arise from imperfect modeling of both the object and the image formation process. While our simulations recreate the most important contrast features, the real measurements have contrast with a broader axial extent and poorer transverse spatial resolution than our simulations---likely due to imperfections in our phantom and slightly aberrated imaging. 

We compared an earlier ray-optics based voxel-by-voxel reconstruction algorithm~\cite{guo_revealing_2020} with WaveOrder's improved wave optical reconstruction algorithm (\textbf{Ext.\,Data\,Fig.\,\ref{extfig:vector_quantitative_reconstruction}c}). We find that wave-optical reconstructions yield marginally improved transverse resolution (\textbf{Ext.\,Data\,Fig.\,\ref{extfig:vector_quantitative_reconstruction}d}) measured via transverse modulation transfer functions from azimuthal profiles, and denoised and defocus-symmetric axial profiles (\textbf{Ext.\,Data\,Fig.\,\ref{extfig:vector_quantitative_reconstruction}e}). We also observe orientation reversals between spokes, reconstruction artifacts that are analogous to well-known ringing artifacts in fluorescence deconvolution.

Together, these examples (\textbf{Figs.\,\ref{fig:quantitative_reconstruction}--\ref{fig:length_scales}}) illustrate that WaveOrder's multi-contrast reconstructions can enhance interpretability of complex biological specimens at scale from single cells to whole organisms. 

\subsection{Democratizing label-free and fluorescence computational imaging}
Computational imaging pipelines for label-free and fluorescence microscopy often require expertise in optics, inverse problems, GPU computation, and workflow engineering. These requirements limit accessibility, particularly for users who need robust reconstructions without deep optics and algorithmic knowledge. WaveOrder lowers these barriers by providing a unified, scalable framework that accommodates exploratory use, routine processing, and advanced customization. 

The key feature that makes WaveOrder scalable is the combination of auto-tuned parameters and tile-based blind deconvolution (\textbf{Fig.\,\ref{fig:tiles}}). This feature is particularly valuable for large-scale optical pooled screens, where the size of the arrays needed to implement deconvolution on the whole stitched image exceeds the RAM available on high-performance computing nodes.

The core of WaveOrder’s accessibility is its multi-interface design. A Napari-based GUI offers an interactive environment for visualization, parameter tuning, and reproducible configuration. For batch and HPC workflows, a configuration-file-driven CLI exposes the same reconstruction parameters, ensuring reproducibility between local and distributed compute environments. Both GUI and CLI read and write OME-Zarr, a standard imaging format with support across the community \cite{moore_ome-zarr_2023}.

For users developing custom methods or integrating WaveOrder into broader pipelines, a modular Python API provides explicit access to forward models, reconstruction operators and optimization routines. The API accepts standard array types (NumPy, Zarr, Dask, PyTorch), enabling efficient use across CPUs, GPUs and cloud infrastructure. Comprehensive documentation, including tutorials and algorithmic detail, is available on \href{https://waveorder.readthedocs.io/en/latest/}{WaveOrder's documentation page (waveorder.readthedocs.io)}. The software engineering effort has enabled broad adoption of WaveOrder by several computational biologists who are not experts in optics or physics-informed machine learning. As a result, WaveOrder has been used to process $\approx1$ PB worth of imaging data across scales from organelles to whole organisms.  

To further broaden access, WaveOrder is available as a browser-based interactive demo on \href{https://huggingface.co/spaces/chanzuckerberg/WaveOrder}{HuggingFace Spaces (huggingface.co/spaces/chanzuckerberg/WaveOrder)}, allowing users to experiment with an example auto-tuned reconstruction workflow without installation or compute resources. A quickstart notebook and model card on the  \href{https://virtualcellmodels.cziscience.com/model/waveorder}{Biohub Virtual Cell platform (virtualcellmodels.cziscience.com/model/waveorder)} provide an immediately runnable environment for onboarding, benchmarking, and reproducible demonstrations.

Together, these interfaces and resources make WaveOrder a flexible and approachable solution for computational imaging across label-free and fluorescence microscopy modalities, supporting users from bench biologists to computational imaging experts.

\section{Discussion}
WaveOrder provides a differentiable, physics-informed machine learning framework for designing multi-contrast imaging systems and for solving shift-variant blind deconvolution problems in label-free and fluorescence light microscopy. By leveraging a unified mathematical representation of single-scattering and diffraction of unpolarized and polarized light, WaveOrder unlocks computational imaging across microscope modalities, contrast types, and biological length scales.

WaveOrder's core strength is its unified forward model. The same framework represents widefield, oblique illumination, oblique detection, confocal, light-sheet, and polarization-resolved microscopes, a sizable fraction of all light microscopy, particularly for live-cell and live-organism imaging. Changing between these modes requires adjusting a small set of physical parameters instead of rewriting reconstruction pipelines for each specific modality, a widespread practice in computational microscopy. 

WaveOrder's parametric representation of microscopes enables its application to diverse biological length scales. The same operators successfully reconstruct data from subcellular organelles, adherent cells, tissues, embryos, and adult zebrafish. WaveOrder assumes linear, single-scattering. Under these assumptions, contrast generation reduces to dipole emission and interference, allowing WaveOrder to treat phase, fluorescence, birefringence, and diattenuation within a single framework. This portability accelerates modeling or optimization across instruments and sample types.

WaveOrder's forward model makes the following key assumptions: 1) The framework assumes channel linearity, meaning each data channel is a linear mixture of underlying material properties. This excludes modeling and reconstruction of data acquired with nonlinear inter-channel phenomena such as FRET, two photon microscopy, and harmonic generation microscopy. 2) WaveOrder also uses weak, single scattering models, which remain accurate for many biological samples but break down in optically thick or heterogeneous samples---adult zebrafish, dense organoids, and thick tissues---where multiple scattering alters both phase and polarization. 3) The framework assumes that the fluorescence image is negligibly affected by the refractive index distribution of the specimen, which can break down in inhomogeneous tissues. Addressing these limitations will require extensions of our forward models with improved treatments of nonlinear interactions, multiple scattering, and mixed contrast modes. 

WaveOrder extends its versatility by pairing parameterized forward models with physics-guided ML to correct shift-variant aberrations. By dividing each acquisition into overlapping, approximately shift-invariant tiles and assigning a small number of unknown parameters to each tile, WaveOrder is able to simultaneously estimate and correct shift-variant aberrations that commonly limit high-throughput imaging screens. Importantly, WaveOrder's optimizations are constrained by analytic transfer functions, keeping the search space low dimensional and physically realizable, preventing the instabilities and non-physical reconstructions that often accompany purely data-driven approaches. In practice, WaveOrder's physics-guided auto-tuning restores consistent contrast across centimeter-scale wells, oblique geometries and multi-channel acquisitions, enabling downstream biological analyses, including segmentation, phenotype quantification, and spatial pattern assessment, to be performed reliably across the entire field of view. 

The key limitations of the current inverse algorithm and software are: 1) WaveOrder currently relies on Tikhonov-regularized reconstructions, which provide simple, efficient, and stable solutions that are not always optimal for low-SNR data, vector reconstructions, or samples with sharp boundaries. Tikhonov regularization implies a Gaussian-distributed noise model, leading to model mismatch in Poisson-noise regimes. More expressive priors, including spatially varying regularizers or plug-and-play schemes, are not yet integrated and will likely extend WaveOrder's value and performance on challenging samples. 2) WaveOrder's auto-tuning workflow requires the user to specify several choices---tile size, tile overlap, parameter initialization, step sizes, and stopping criteria---that all affect performance and accessibility. While we find our physics-guided optimizations to be quite robust, WaveOrder does not alert the user when auto-tuning fails. These limitations point toward future developments in adaptive regularization, noise modeling, and automated configuration to improve robustness across diverse use cases.  

WaveOrder's modeling choices balance physical accuracy with practical usability. We adopt a vector wave-optical formulation because interference, diffraction, and polarization govern the contrast mechanisms most widely used in biology, fluorescence, phase, and linear anisotropy, where ray-optics models are inadequate. At the same time we restrict our forward model to single scattering, which captures the dominant behavior in thin and moderately thick specimens while avoiding the instability, computational burden, and extreme non-invertability of full multiple-scattering treatments. Embedding these physical constraints into the reconstruction operation sharply reduces the size of the blind deconvolution search space, eliminating the need to train large neural networks to approximate optical physics or to rely on simulation-heavy supervision. WaveOrder's physics-first design yields interpretable reconstructions, prevents hallucinated structure, and provides a foundation on which more flexible learned priors can be incorporated into future work.  

WaveOrder not only improves contrast, but also unlocks quantitative biological measurements. Because WaveOrder reconstructs physically interpretable parameters such as phase, birefringence, and fluorescence density, these measurements map cleanly onto underlying biophysical processes rather than arbitrary image intensities, enabling more reliable and mechanistic biological conclusions. Reconstructions substantially increase the quality of downstream segmentation and quantification, illustrated by gain in cell segmentation performance in pooled screens and in the recovery of fine organelle structures that are nearly invisible in raw data. Similar improvements appear in cell type classification tasks: neuromast cell classes that overlap in raw fluorescence become separate after reconstruction, and multispectral iPSC data show improved correlations between phase and fluorescence channels, revealing relationships obscured by optical blur. The framework also resolves sarcomere periodicity in cardiomyocytes, vacuole-size gradients along the zebrafish notochord, and density variations in the zebrafish retina.  

WaveOrder is designed to be deployed across a wide range of computational environments, lowering practical barriers to applying physics-based reconstruction methods. A Napari GUI supports interactive exploration and parameter tuning, while a configuration-driven CLI enables batch processing on HPC clusters. Both GUI and CLI read and write OME-Zarr, ensuring compatibility with community tools. For rapid onboarding, a browser based demo and notebook provide interactive examples without installation. At the developer level, a modular Python API exposes forward models, transfer functions, and reconstruction operators, allowing integrations into larger imaging pipelines and custom development. Together, these components form a coherent software suite that makes WaveOrder accessible to non-experts and extensible for advanced users.  

Several extensions could broaden WaveOrder's capabilities. Support for multiple scattering, using multislice and beam-propagation models \cite{lim_beyond_2017, chowdhury_high-resolution_2019, mu_multislice_2023, bartels_analysis_2024}, would make the framework applicable to thicker and more heterogeneous samples. WaveOrder can also act as a physics-constrained front end for neural networks; earlier work showed that supplying physically corrected inputs improves downstream virtual staining performance \cite{liu_robust_2025}, and we expect that including an inverted physical forward model as a differentiable first layer should generalize this benefit to segmentation \cite{stringer_cellpose_2021, pachitariu_cellpose_2022}, tracking \cite{bragantini_ultrack_2025, gallusser_trackastra_2025}, phenotyping, and forecasting tasks \cite{hirata-miyasaki_dynaclr_2024, shannon_cellplatounsupervised_2024}. Finally, the same parameter estimates used here for physics-guided post-acquisition reconstruction could drive smart microscopy workflows \cite{carpenter_smart_2023, marin_navigate_2024}, enabling shift-variant acquisitions with real-time adaptive-optical modifications of illumination and detection paths. These directions push toward tighter integration between modeling, learning, and microscope control. 

To conclude, we find that linear models provide a strong framework for widely applicable computational microscopy techniques, including phase, absorption, birefringence, diattenuation, and anisotropic fluorescence imaging. We find the WaveOrder framework useful for understanding, simulating, and reconstructing data acquired with this class of techniques, and we demonstrate its ability to improve multi-contrast multi-channel data across length scales.

\end{refsegment}

\clearpage
\printbibliography[segment=1,title={References}]
\clearpage

\begin{refsegment}
\section{Methods}\label{sec:methods}
\subsection{Sample preparation and imaging}\label{sec:sample_prep}
Sample preparation and imaging methods are described in \textbf{Supp.\,8}. 

\subsection{WaveOrder framework}\label{sec:framework}
We start by describing a framework for reconstructing material properties from microscopic imaging data. By representing material properties and data as vectors, we describe our reconstructions as the solution to an optimization problem. Using broadly applicable assumptions of linearity, shift invariance, and weak scattering, we describe all contrasts as the result of banks of point spread functions. We describe the physics behind the major contrast mechanisms and compute their transfer functions from illumination, scattering, and detection models. Throughout, we use carefully chosen basis functions to enable the physical interpretation of light and material properties. 

We describe all notation as we introduce it, and we collect all symbols in \textbf{Supp.\,1}.

\subsubsection{Objects and data as vectors}\label{sec:objects_and_data}
We represent a biological sample as a series of \textit{volumetric maps of material properties}, shown schematically in \textbf{Fig.\,\ref{fig:overview}c,\,i}. For example, a cell might be approximately described by two fluorophore density maps, $f_1(\rvec_o)$ and $f_2(\rvec_o)$ where $\rvec_o$ is a 3D object position vector, one map for each of two different fluorophores that label biological structures of interest, and a density map $\rho(\rvec_o)$. Three maps are unlikely to describe the sample completely, so we generalize and collect any number of volumetric maps into a single vector
\begin{align}
    \mathbf{f} = [\mathbf{f}_1, \mathbf{f}_2, \mathbf{f}_3, \hdots]^T = [f_1(\rvec_o), f_2(\rvec_o), \rho(\rvec_o), \hdots]^T,
\end{align}
which represents all of the properties of our sample. 

When we image our sample in a microscope, we arrange for the material properties to be encoded into a list of \textit{volumetric datasets}, each called a \textit{channel}, shown schematically in \textbf{Fig.\,\ref{fig:overview}c,\,ii}. For example, we might use a fluorescence light path to encode fluorophore density maps into two channels, $d_1(\rvec_d)$ and $d_2(\rvec_d)$ where $\rvec_d$ is a 3D detector position vector, then we can change to a transmission light path to encode the density $\rho(\rvec_o)$ into the third channel $d_3(\rvec_d)$. Similar to our object properties, we collect any number of channels into a single vector
\begin{align}
    \mathbf{d} = [\mathbf{d}_1, \mathbf{d}_2, \mathbf{d}_3, \hdots]^T = [d_1(\rvec_d), d_2(\rvec_d), d_3(\rvec_d), \hdots]^T,
\end{align}
which represents all of the data we collect from our sample. 

\subsubsection{Imaging and reconstruction as linear operators}\label{sec:linear_operators}
We can represent the imaging process with a single \textit{forward operator} $\mathcal{H}$ that encodes material properties $\mathbf{f}$ into measured volumetric datasets $\mathbf{d}$
\begin{align}
    \mathbf{d} = \mathcal{H}\mathbf{f} + \mathbf{b},\label{eq:fwd}
\end{align}
where $\mathbf{b}$ is a spatially uniform background in each channel. Note that $\mathcal{H}$ might encode multiple material properties into a single channel. For example, the material properties of phase and anisotropy can be jointly encoded into several label-free data channels.

\subsubsection{Reconstructing object properties}\label{sec:reconstructing_high}
We would like to recover as much as we can about the object's material properties $\mathbf{f}$ from the measured data $\mathbf{d}$, but we are faced with a major problem---\textit{the forward operator $\mathcal{H}$ is never invertible}. There are always object properties that are invisible to the imaging system, and one way to find invisible properties is to make the properties smaller than the resolution limit of the imaging system. For example, if we have a visible-light microscopy dataset $\mathbf{d}$ there are an infinite number of molecular-scale configurations that could result in the same dataset, so we have no hope of choosing a single $\mathbf{f}$ as \textit{the} true measured properties. 

We need to choose a single set of material properties from among the infinite possible solutions that agree with the data---a \textit{reconstruction problem}. Our strategy is to choose the material properties that minimize a scalar objective function $Q(\mathbf{f}, \mathbf{d})$ 
\begin{align}
    \mathbf{\hat{f}} = \mathcal{R}\mathbf{d} = \argmin_{\mathbf{f}} Q(\mathbf{f}, \mathbf{d}),
\end{align}
where the $\argmin$ notation means that we choose as our solution $\mathbf{\hat{f}}$ the $\mathbf{f}$ that minimizes the value of $Q(\mathbf{f}, \mathbf{d})$. One choice is the \textit{least-squares objective} $Q^{\text{(ls)}}(\mathbf{f}, \mathbf{d}) = \left\|\mathbf{d} - \mathbf{b} - \mathcal{H}\mathbf{f}\right\|_2^2,$ but this solution tends to amplify noise. A better choice is a \textit{Tikhonov-regularized least-squares objective}  
\begin{align}
    Q_\eta^{\text{(tls)}}(\mathbf{f}, \mathbf{d}) = \left\|\mathbf{d} - \mathbf{b} - \mathcal{H}\mathbf{f}\right\|_2^2 + \eta \|\mathbf{f}\|_2^2,
\end{align}
which adds a regularization parameter $\eta$ that suppresses the size of the solution $\mathbf{f}$, an example of a \textit{prior} that penalizes large solutions. Many other objective functions are possible, including those that include physics-informed and learned priors. 

After choosing an objective function, the minimization problem needs to be solved, an often challenging task. Fortunately, if $\mathcal{H}$ is linear then the Tikhonov-regularized least-squares objective can be minimized in a single step (\textbf{Methods\,\ref{sec:recon_low}}), generating efficient noise-tolerant estimates of material properties (\textbf{Fig.\,\ref{fig:overview}c,\,iii}).

\subsubsection{Label-free and fluorescence contrast in a unified framework}\label{sec:unified}
All contrast is formed by illuminating the sample with electric fields that scatter from the sample then interfere on the detector. If we consider only single scattering events, the \textit{first Born approximation}, then we can rewrite \textbf{Eq.\,\ref{eq:fwd}} as 
\begin{align}
    \mathbf{d} = \left|\mathcal{S}\mathbf{f} + \mathbf{e}^{\text{(d)}}\right|^2,
\end{align}
where $\mathcal{S}$ is a \textit{scattering operator} that models the scattered fields that reach the detector and $\mathbf{e}^{(d)}$ models the unscattered \textit{direct fields} that reach the detector. Expanding the square reveals four terms 
\begin{align}
    \mathbf{d} = |\mathcal{S}|^2|\mathbf{f}|^2 + \mathbf{e}^{\dagger\text{(d)}}\mathcal{S}\mathbf{f} + \mathbf{f}^{\dagger}\mathcal{S}^{\dagger}\mathbf{e}^{(\text{d})} + |\mathbf{e}^{\text{(d)}}|^2,
\end{align}
that we refer to as the scatter-scatter, scatter-direct, direct-scatter, and direct-direct terms, respectively, and $\dagger$ denotes conjugate transpose. 

We consider two classes of contrast. \textit{Label-free contrast} (\textbf{Fig.\,\ref{fig:overview}a,\,iii}) is generated by illuminating the sample with light that interacts with the sample \textit{coherently}---that is, scattered fields have the same wavelength and a fixed phase relationship with the illuminating fields. When a plane wave encounters a coherent scatterer, the oscillating electric field accelerates bound electrons in the scatterer, and these accelerated charges generate spherical scattered fields. The direct and scattered fields interfere and generate contrast via the scatter-direct and direct-scatter terms. The direct-direct term creates a uniform background, and for \textit{weakly scattering} samples the scatter-scatter term is small and ignorable. Therefore, label-free contrast is generated by the direct-scatter and scatter-direct terms on top of a direct-direct background. Finally, each point on the source emits incoherently, so we can treat each source point individually and find the complete contrast pattern by \textit{summing over the source}. 

\textit{Fluorescence contrast} (\textbf{Fig.\,\ref{fig:overview}a,\,iv}) is generated by illuminating fluorescent scatterers and imaging their scattered light. Fluorescent scatterers are \textit{incoherent}, so the scattered fields have a random phase at a longer wavelength than the illuminating fields. Therefore, the scatter-direct and direct-scatter terms do not generate contrast, so the only way to measure sample-dependent contrast is via the small scatter-scatter term. Fortunately, the direct and scattered fields are at different wavelengths, so the direct fields can be filtered with minimal bleedthrough. Therefore, fluorescence contrast is generated by the scatter-scatter term with a direct-direct bleedthrough background. Finally, fluorescent scatterers emit incoherently, so we can find the complete contrast pattern by \textit{summing over the sample}.

Both label-free and fluorescence contrast modes can generate additional contrast from \textit{anisotropic samples}. Label-free samples can be anisotropic if the scatterer's bound electrons accelerate anisotropically. We illustrate a label-free anisotropic sample schematically as an electron bound to its nucleus by springs of varying spring constant (\textbf{Fig.\,\ref{fig:overview}a,\,v}). When polarized light is incident on an anisotropic sample, it accelerates the bound electrons in linear, circular, or elliptical \textit{dipoles}, which emit anisotropic polarized light in patterns that encode the orientation of the induced electron motion and the underlying anisotropy of the scatterer. Therefore, information about the sample's label-free anisotropy is encoded in the polarization and intensity pattern of the detected light. Similarly, fluorescent scatterers emit along linear, circular, or elliptical dipoles (\textbf{Fig.\,\ref{fig:overview}a, vi}), though linear dipoles are most common among the fluorophores used in biological microscopy.

\subsubsection{Physically interpretable basis functions}\label{sec:interpretable_basis}
When we illuminate a label-free sample, the 3D induced dipole moment is the product of the incident field and a $3\times 3$ matrix called the \textit{permittivity tensor}~\cite{yeh_permittivity_2024, wolf_introduction_2007}. By convention, we change to a unitless quantity and subtract the isotropic background (\textbf{Supp.\,3}) to arrive at a complete set of label-free sample properties---the complex-valued 3 $\times$ 3 matrix called the \textit{scattering potential tensor}, $f^{(\text{lf})}_{ij}$. Each entry of the scattering potential tensor can be interpreted directly (e.g.\,the complex-valued $f^{(\text{lf})}_{xz}$ is the relative magnitude and phase of the $x$ component of the dipole induced by a $z$-oriented field), but this interpretation can be challenging to understand physically. To improve physical interpretability, we expand the scattering potential tensor onto the \textit{spherical harmonic tensors}, a set of nine 3 $\times$ 3 matrices whose complex-valued expansion coefficients can be directly interpreted in terms of phase, absorption, birefringence, and diattenuation. We schematize each of these spherical harmonic tensors in \textbf{Ext.\,Data\,Fig.\,\ref{extfig:basis-functions}a} by drawing each tensor's eigenvalues and eigenvectors, and we describe the spherical harmonic tensor basis in detail in \textbf{Supp.\,3}.

In a fluorescent sample, the 3D emission dipole moment can be represented by a three-component vector $f^{(\text{fl})}_{i}$ (\textbf{Fig.\,\ref{fig:overview}a,\,vi;\,Ext.\,Data\,Fig.\,\ref{extfig:basis-functions}b}) with real-valued coefficients for purely linear dipoles and complex-valued coefficients for arbitrary dipoles. Contrast arises from the scatter-scatter term, so our measurements are proportional to the squares of the dipole components $|f^{(\text{fl})}_{i}|^2$. 

For dynamic ensembles of fluorescent emitters, the measurements are proportional to the \textit{second-moment matrix} $\langle f^{(\text{fl})}_{i}f^{*(\text{fl})}_{j} \rangle$~\cite{backer_extending_2014, zhang_six-dimensional_2023, zhang_single-molecule_2021}. Similar to the scattering potential tensor, we can expand the second-moment matrix onto the spherical harmonic tensors, but here we interpret the coefficients in terms of orientation distribution functions~\cite{chandler_spatio-angular_2019, chandler_volumetric_2025}. 

Finally, we express our data in terms of the \textit{Stokes parameters}, a set of four real-valued parameters that are physically interpretable as the intensities measured behind various polarizing filters. The Stokes parameters can also be interpreted as the coefficients of the electric field's second-moment matrix expanded onto the \textit{Pauli matrices} (\textbf{Ext.\,Data\,Fig.\,\ref{extfig:basis-functions}c}, \textbf{Supp.\,4}).

{
\setlength{\tabcolsep}{12pt}
\renewcommand{\arraystretch}{1.0}
\begin{table*}
\centering
\scriptsize
\begin{tabular}{r|cc}
 & vector & scalar\\
\hline\\
label-free & $\begin{bmatrix}H^{(\text{lf,re})}_{cp}(\nuvec)\\ H^{(\text{lf,im})}_{cp}(\nuvec)\end{bmatrix} = \sigma_c^{ii'}\begin{bmatrix}
    1 & 1\\
    \iim & - \iim
\end{bmatrix}
\begin{bmatrix}
[P_{ij}G_{jn}\starr P_{i'k'}S_{k'}S^*_{k}](\nuvec)\\
[P_{ik'}S_{k'}S^*_{k}\starr P_{i'j}G_{jn}](\nuvec)
\end{bmatrix}\mathcal{Y}_{p}^{nk}$ & $\begin{bmatrix}H^{(\text{lf,re,s})}(\nuvec)\\ H^{(\text{lf,im,s})}(\nuvec)\end{bmatrix} = \begin{bmatrix}
    1 & 1\\
    \iim & - \iim
\end{bmatrix}
\begin{bmatrix}
    [P \star PS](\nuvec)\\
    [PS \star P](\nuvec)
\end{bmatrix}$\\ \\
fluorescence & $H^{(\text{fl})}_{cp}(\nuvec) = \sigma_c^{ii'}[P_{ij}G_{jk}\star P_{i'j'}G_{j'k'}](\nuvec)\mathcal{Y}_{p}^{kk'}$ & $H^{(\text{fl,s})}(\nuvec) = [P\star P](\nuvec)$ \\
\end{tabular}
\caption{\textbf{WaveOrder's transfer functions in terms of autocorrelated pupils.} Transfer functions for different contrast modes (rows) and optical models (columns) expressed in terms of the illumination model $S$, the scattering model $G$, the detection model $P$, the object-space basis $\mathcal{Y}$, and the data-space basis $\sigma$. Sums over $i, i', j, j', k$ and $k'$ are implied.}\label{tab:tf}
\end{table*}
}

\subsubsection{Contrast-separable imaging systems}\label{sec:contrast_separable}
In many microscopy imaging systems, data in each channel is the sum of the contributions from each material property. We call these systems \textit{channel linear} and we can express the forward operator as
\begin{align}
    \mathbf{d}_c = \mathcal{H}_{cp}\mathbf{f}_{p} + \mathbf{b}_c
\end{align}
where $c$ indexes channels and $p$ indexes material properties. For example, multi-channel fluorescence microscopy often suffers from crosstalk, but these systems are still channel linear if the data is the sum of the contribution from each type of fluorophore.    

When there is no crosstalk between channels (for example, when fluorescence filters are perfect), we say the imaging system is \textit{channel separable}, which implies that the system can be written as   
\begin{align}
    \mathbf{d}_c = \mathcal{H}_{cc}\mathbf{f}_c + \mathbf{b}_c,
\end{align}
i.e. the operator $\mathcal{H}$ is \textit{diagonal over channels}.

The WaveOrder framework considers imaging systems where some groups of channels are separable (e.g. fluorescence contrast), and some groups of channels are merely linear (e.g. label-free contrast). We call such imaging systems \textit{contrast separable} and they can be expressed as
\begin{align}
    \mathbf{d}^{(m)}_c = \mathcal{H}^{(m)}_{cp}\mathbf{f}^{(m)}_{p} + \mathbf{b}_c^{\text(m)},\label{eq:fwd_block}
\end{align}
where $m$ indexes each of the $M$ \textit{contrast modes}. For example, an imaging system with two channel-separable fluorescent channels and four channel-linear label-free data channels that jointly encode three object properties can be written as
\begin{align}
\begin{bmatrix} \mathbf{d}_1^{(1)} \\ \mathbf{d}_1^{(2)} \\ \mathbf{d}_1^{(3)} \\ \mathbf{d}_2^{(3)} \\ \mathbf{d}_2^{(3)} \\ \mathbf{d}_3^{(3)}\end{bmatrix} = 
\setlength{\arraycolsep}{2pt}
\begin{bmatrix} 
\mathcal{H}^{(1)}_{1,1} & 0 & 0 & 0 & 0\\ 
0 & \mathcal{H}^{(2)}_{1,1} & 0 & 0 & 0\\ 
0 & 0 & \mathcal{H}^{(3)}_{1,1} & \mathcal{H}^{(3)}_{1,2} & \mathcal{H}^{(3)}_{1,3}\\ 
0 & 0 & \mathcal{H}^{(3)}_{2,1} & \mathcal{H}^{(3)}_{2,2} & \mathcal{H}^{(3)}_{2,3}\\ 
0 & 0 & \mathcal{H}^{(3)}_{3,1} & \mathcal{H}^{(3)}_{3,2} & \mathcal{H}^{(3)}_{3,3}\\ 
0 & 0 & \mathcal{H}^{(3)}_{4,1} & \mathcal{H}^{(3)}_{4,2} & \mathcal{H}^{(3)}_{4,3} 
\end{bmatrix}
\begin{bmatrix} \mathbf{f}^{(1)}_1 \\ \mathbf{f}^{(2)}_1 \\ \mathbf{f}^{(3)}_1\\ \mathbf{f}^{(3)}_2 \\ \mathbf{f}^{(3)}_3\end{bmatrix} + 
\begin{bmatrix} \mathbf{b}_1^{(1)} \\ \mathbf{b}_1^{(2)} \\ \mathbf{b}_1^{(3)} \\ \mathbf{b}_2^{(3)} \\ \mathbf{b}_2^{(3)} \\ \mathbf{b}_3^{(3)}\end{bmatrix}.
\end{align}
In other words, we assume that the forward operator is \textit{block diagonal over channels}, so we can split our reconstruction problem into subproblems, one for each contrast mode
\begin{align}
    \mathbf{\hat{f}}_{p}^{(m)} = \mathcal{R}_{\eta}^{(m)}\mathbf{d}_c^{(m)} = \argmin_{\mathbf{f}_{p}^{(m)}} Q_{\eta}^{\text{(tls)}}\left(\mathbf{f}_{p}^{(m)}, \mathbf{d}_c^{(m)}\right),\label{eq:recon}
\end{align}
where we have defined $\mathcal{R}_{\eta,pc}^{(m)}$ as a \textit{reconstruction operator} for each contrast mode. \textbf{Eq.\,\ref{eq:recon}} is the core of our reconstruction algorithm. 

\subsubsection{Linear contrast-separable shift-invariant imaging systems}
When a contrast-separable imaging system is approximately spatially linear and shift invariant, we can express our imaging model as      
\begin{align}
d^{(m)}_c(\rvec_d) = \sum_{p} \int d\rvec_o\, h^{(m)}_{cp}(\rvec_d - \rvec_o)f^{(m)}_{p}(\rvec_o) + b_c^{(m)}, 
\end{align}
where $m$ indexes contrast $m$odes, $p$ indexes material $p$roperties, $c$ indexes data $c$hannels, and $h^{(m)}_{cp}$ is a bank of \textit{point spread functions} that model the entire multi-contrast multi-channel imaging system. We can reexpress this relationship in the Fourier domain as 
\begin{align}
    D^{(m)}_c(\nuvec) = \sum_{p} H^{(m)}_{cp}(\nuvec)F^{(m)}_{p}(\nuvec) + b_c^{(m)}\delta(\nuvec).
\end{align}
where $\nuvec$ is a 3D spatial frequency coordinate, capital letters denote 3D Fourier transforms, and $H^{(m)}_{cp}(\nuvec)$ is a bank of \textit{transfer functions} that model the transmission of spatial frequency components through the imaging system. We inspect the properties of these transfer functions next.  

\subsubsection{Summary of transfer functions}\label{sec:tf_summary}
The WaveOrder framework calculates all transfer functions from three core submodels: 
\begin{enumerate}\itemsep0em
    \item an illumination model---the vector source pupil $S_i(\nuvec)$,
    \item a scattering model---the Green's tensor spectrum $G_{ij}(\nuvec)$,
    \item a detection model---the tensor detection pupil $P_{ij}(\nuvec)$. 
\end{enumerate} 
All three submodels are expressed as complex-valued spherical shell functions with radius $1/\lambda$ in the frequency domain, where $\lambda$ is the wavelength in the imaging media. 

The Green's tensor spectrum is particularly important for modeling anisotropic contrast. Linear dipole moments emit polarized light in a doughnut-shaped intensity pattern (\textbf{Ext.\,Data\,Fig.\,\ref{extfig:transfer}a}), and the Green's tensor spectrum (\textbf{Ext.\,Data\,Fig.\,\ref{extfig:transfer}b}) efficiently models all dipole emitters (coherent or incoherent; linear, circular, or elliptical dipoles in any orientation) with a single function. 

All transfer functions can be expressed as products and autocorrelations of the illumination, scattering, and detection models, see \textbf{Table \ref{tab:tf}} and \textbf{Supp.\,5}. We refer to the complete transfer functions as \textit{vector models} because they account for the complete vectorial nature of light and dipole scattering, and we also include scalar models that ignore vector effects, which are reasonable approximations when unpolarized illumination and unpolarized detection are used on isotropic samples.

\textbf{Ext.\,Data\,Figs.\,\ref{extfig:transfer}c--f} show the support and phase of several examples of WaveOrder's transfer functions. We briefly highlight several key features
\begin{itemize}\itemsep0em
    \item vector models consist of a grid of transfer functions, one for each data channel and material property,
    \item scalar models consist of a single transfer function, and
    \item we model real-valued data, so our transfer functions are \textit{Hermitian}, that is $H_{cp}^{(m)}(\nuvec) = H_{cp}^{*(m)}(-\nuvec)$.
\end{itemize}

\subsubsection{Reconstruction algorithm}\label{sec:recon_low}
For each contrast mode, the forward operators in \textbf{Eq.\,\ref{eq:fwd_block}} can be decomposed using the singular value decomposition
\begin{align}
\mathcal{H} = \sum_i^R s_i \mathbf{u}_{i}\mathbf{v}^{\dagger}_{i},
\end{align}
where $R$ is the rank of $\mathcal{H}$, $\{\mathbf{u}_{i}\}$ is a set of orthonormal data-space vectors that span the space of deterministic expected data, $\{\mathbf{v}_{i}\}$ is a set of orthonormal object-space vectors that span the measurement space of object properties, and $\{s_i\}$ are singular values.

We use this decomposition to solve \textbf{Eq.\,\ref{eq:recon}} in a single step for each mode
\begin{align}
\mathcal{R}_{\eta} = \sum_i^R \frac{s_i}{s_i^2 + \eta}\mathbf{v}_{i}\mathbf{u}^{\dagger}_{i}.\label{eq:tls_solution}
\end{align}
In practice we choose a different regularization parameter $\eta$ for each contrast mode. 

\subsection{Cell-scale optical pooled screen data}\label{sec:ops}
\textbf{Auto-tuned reconstructions} 2D phase reconstructions were computed from multi-plane volumetric data using a phase-from-defocus forward model. For each field of view, the volumetric data were divided into a $6\times 6$ grid of tiles with 20\% overlap.

To optimize reconstruction quality, we implemented gradient-based auto-tuning of the model parameters. The objective function maximized the power spectral density in a mid-band spatial frequency range (1/8--1/4 of the transverse cutoff frequency), which corresponds to cellular feature sizes while avoiding low-frequency background variations and high-frequency noise. Three parameters were optimized: axial focal offset ($z_{\text{offset}}$), illumination zenith tilt angle ($\theta_{\text{zenith}}$), and azimuthal tilt angle ($\phi_{\text{azimuth}}$).

The optimization used the Adam algorithm~\cite{kingma_adam_2014} with learning rates of 0.02 \textmu m, 0.005 rad, and 0.001 rad per iteration for $z_{\text{offset}}$, $\theta_{\text{zenith}}$, and $\phi_{\text{azimuth}}$, respectively. Each tile was optimized for 100 iterations, with the reconstruction and gradient computation performed in PyTorch~\cite{paszke_pytorch_2019} to enable automatic differentiation. The regularization strength for the inverse problem was fixed at $\eta = 10^{-2}$. Fixed optical parameters included illumination wavelength (450 nm), detection NA (0.15), illumination NA (0.1), and sample refractive index (1.0 for air-mounted samples). Optimization progress was monitored using TensorBoard, logging parameter values, loss curves, and reconstructed phase images at each iteration.

\textbf{Cell segmentation}
Cell segmentation was performed on reconstructed phase images to quantitatively compare reconstruction methods. Three conditions were evaluated: (i) raw multi-plane intensity data from the highest-contrast slice $z$ plane, (ii) nominal reconstruction with zero illumination tilt, and (iii) auto-tuned reconstruction with optimized illumination parameters. All images were cropped to a standardized $300\times 300$ pixel region for analysis.

Segmentation was performed using Cellpose \texttt{cyto2} model~\cite{stringer_cellpose_2021, pachitariu_cellpose_2022}. Cell diameters were set to 175 pixels across all conditions and fields of view, with a flow threshold of 0.9. Ground truth annotations were manually created for four representative image regions and saved as labeled arrays.

Segmentation accuracy was quantified using intersection-over-union (IoU) metrics with Hungarian matching. For each image pair (predicted and ground truth), we computed the IoU between all object pairs and performed optimal one-to-one matching using linear sum assignment to maximize total IoU. This yielded a distribution of per-object IoU scores for each condition.

We computed recall (fraction of ground truth cells successfully detected) and precision (fraction of predicted cells matching ground truth) as functions of IoU threshold $t \in [0, 1]$, sampled at 0.01 intervals. F1 scores were calculated as $\text{F1}(t) = 2 \cdot \text{precision}(t) \cdot \text{recall}(t) / (\text{precision}(t) + \text{recall}(t))$ at each threshold. Precision-recall-F1 curves were generated for each field of view to characterize segmentation performance across the three reconstruction conditions.

\subsection{Organelle-scale optical pooled screen data}\label{sec:organelle}
\textbf{Tiling and stitching:}
Each field of view was divided into a $5\times 5$ grid of tiles with 25 pixels of overlap. After reconstruction, the tiles were stitched with linearly weighted Euclidean distance transform (EDT)-based blending. 

\textbf{Auto-tuned reconstructions:}
 Raw brightfield through-focus stacks were processed using a scalar reconstruction routine to generate 3D phase density maps. Additionally, a phase-from-defocus model was auto-tuned by optimizing the axial focal offset ($z_{\text{offset}}$) for power spectral density in a mid-band spatial frequency range (1/8 --1/4 of the transverse cutoff frequency), summarizing the cellular phenotype into a single high-SNR plane.  

\textbf{Image registration:}
To account for chromatic aberrations and mechanical shifts between fluorescence (ground truth) and label-free channels, we implemented an automated registration routine. For every field of view, a translational offset sweep ($\pm 5$ pixels in $X$ and $Y$) was performed. The optimal offset was determined by maximizing the spatial overlap between the Frangi-filtered binary masks of the fluorescence channel and the label-free reconstruction. This offset was applied prior to feature extraction to ensure accurate spatial comparison.

\textbf{Organelle segmentation:}
Organelles were segmented using a multi-scale Frangi vesselness filter adapted for distinct morphologies~\cite{lefebvre_nellie_2025}. Prior to filtering, label-free images (raw, 3D reconstruction, 2D reconstruction) were pre-processed with Contrast Limited Adaptive Histogram Equalization (CLAHE; clip limit = 0.01, kernel size = $64 \times 64$ pixels) to normalize local contrast.

\textit{Mitochondria:}
To capture tubular networks, we utilized a Frangi filter with high plate-sensitivity ($\alpha = 4.0$, $\beta = 0.5$) across scales corresponding to radii of $0.1$--$1.5$ \textmu m.

\textit{Endosomes:} 
To capture vesicular/blob-like structures, we utilized a filter with isotropic sensitivity ($\alpha = 0.5$, $\beta = 0.5$) across radii of $0.2$--$1.5$ \textmu m.

Probability maps were thresholded using a combined Triangle/Otsu method on the logarithmic vesselness response~\cite{lefebvre_nellie_2025}. A cytoplasm mask was generated by eroding the cell segmentation mask (2 iterations) and excluding the nucleus (dilated by 20 pixels) to restrict analysis to the cytoplasm and remove nuclear artifacts.

\textbf{Quantitative comparison:}
Segmentation performance was quantified using a spatial overlap metric, defined as the percentage of the binarized fluorescent ground truth area that intersected with the label-free segmentation mask. 

\textbf{Statistical analysis:}  
Statistical comparisons between imaging modalities (raw, 3D reconstruction, 2D reconstruction) were performed treating individual cells as biological replicates. We used a repeated measures ANOVA to test for differences across reconstruction methods, followed by post-hoc paired $t$-tests with Bonferroni correction for pairwise comparisons.

\subsection{Neuromast segmentation and classification}\label{sec:neuromast}
\textbf{Segmentation and annotation:}
Neuromast nuclei were manually segmented using signal from the she:H2B-EGFP transgenic line~\cite{peloggia_adaptive_2021}. Each nucleus in a neuromast was assigned a unique identifier. Nuclei were assigned to one of two categories. The first category, mantle cells, form the boundary of each neuromast. The second category, hair and support cells, contains the internal proliferative and mechanosensory cells of the neuromast.

\textbf{Texture metric calculation:}
Single cell Shannon Entropy values were calculated using the neuromast nuclear instance segmentations described above \cite{shannon_mathematical_1948}. Similarly, homogeneity and correlation values were calculated to have second order metrics describing fluorescent signal texture(s) \cite{haralick_textural_1973}.

\textbf{Statistical analysis:}
Kruskal-Wallis H-tests for independent samples were computed to compare first and second order texture measurements between imaging modalities and reconstructions \cite{kruskal_use_1952}, with significance levels (*) $p<0.05$, (**) $p<0.01$, (***) $p<0.001$.

\subsection{Cardiomyocyte reconstructions and profiles}\label{sec:cardio}
Cardiomyocytes were imaged with polarization-resolved label-free microscopy with 8 sector-shaped illumination patterns and a single full-aperture illumination pattern~\cite{yeh_permittivity_2024}. Raw intensity measurements acquired with four polarization states were averaged to reduce polarization-dependent artifacts, then reconstructed into 3D phase volumes. Two reconstruction strategies were compared: phase reconstruction from a circular illumination aperture versus phase reconstruction combining all nine illumination apertures. Manually drawn line profiles across cellular structures were extracted to quantitatively compare image quality between reconstruction approaches.

\subsection{Zebrafish embryo analysis}\label{sec:zfishembryo}
\textbf{Preprocessing:} 
Raw volumes were deskewed and registered by estimating affine transformations for each view to correct the oblique geometry and align all views into a common coordinate system.

\textbf{Tail:}
\textit{Straightening and profiles:} The zebrafish tail was straightened into rectilinear coordinates using curvilinear resampling. A manually-annotated midline was smoothed with a cubic spline and parameterized by arc length $s$. At each position along the midline, we computed the unit tangent vector $\mathbf{t}(s)$ and perpendicular normal vector $\mathbf{n}(s)$. For the label-free  and nuclei channels, we sampled symmetrically in both directions (dorsal and ventral) from the midline. For membrane fluorescence, we sampled only the dorsal side to isolate the notochord region. Local perpendicular widths were computed adaptively by ray-casting along normals until reaching the mask boundary, and the straightened coordinate system was normalized such that $n \in [0, 1]$ spans from midline to dorsal edge. Image intensities were resampled onto a regular grid in curvilinear $(s, n)$ space using bilinear interpolation.

\textit{Spatial distribution quantification:} The straightened image was divided into anterior and posterior halves along the midline axis. For each half, we computed the total integrated intensity by summing all pixels. The spatial enrichment ratio was calculated as $R = I_{\text{anterior}} / I_{\text{posterior}}$.

\textit{Statistics:} Standard errors for the enrichment ratio were estimated via bootstrap resampling ($n = 10{,}000$ iterations). For each bootstrap sample, we resampled rows (perpendicular to the midline, i.e., along the dorsal axis) with replacement, recomputed the integrated intensities for each half, and calculated the ratio. The standard error of the mean was computed as SEM = $\sigma_{\text{boot}} / \sqrt{n}$, where $\sigma_{\text{boot}}$ is the standard deviation of the bootstrap distribution.

\textbf{Notochord:} \textit{ROI selection and extraction:} Three sub-ROIs (posterior, median, and anterior) were manually selected from the tail region, and rotated rectangular regions were resampled onto a regular grid using bilinear interpolation.

\textit{Profile extraction:} Intensity profiles were extracted along the long axis (horizontal center) of each straightened sub-ROI. Profiles were computed over the central region spanning 10--40\% of the ROI length, averaging intensities across 5 pixels perpendicular to the profile direction. For each profile line, the sum of intensities across the 5-pixel width was computed, yielding a 1D profile representing integrated signal along the notochord axis.

\textit{Peak detection and spacing quantification:} Peaks in the phase reconstruction profiles were manually identified from the normalized intensity traces, and distances between consecutive peaks were computed.

\textit{Statistics:} For each region (posterior, median, anterior), mean peak spacing was calculated with standard error of the mean (SEM = $\sigma / \sqrt{n}$) and standard deviation (SD). Statistical comparisons between regions were performed using the Mann-Whitney U test (two-sided) with significance levels (*) $p<0.05$, (**) $p<0.01$, (***) $p<0.001$.

\textbf{Eye:}
\textit{Annotation and resampling:} Similar to the tail, we manually annotated the retina boundary and midline, resampled onto a Cartesian grid, then plotted 1D projections along anatomical axes. 

\textit{Profiles:} The resampled retina was divided into basal, medial, and apical thirds. For each zone and fluorescence channel, we computed total integrated intensity $I_{\text{zone}}$ and normalized to the medial zone: $I_{\text{zone}}^{\text{rel}} = I_{\text{zone}} / I_{\text{medial}}$.

\textit{Statistics:} Standard errors were estimated by computing per-column mean intensities within each zone, normalizing to the medial intensity, and calculating SEM = $\sigma / \sqrt{M}$ across $M$ valid columns.

\section{Extended data}
\newcounter{exttab}
\begin{sidewaystable*}[tp]
\centering
\small
\setlength{\tabcolsep}{4pt}
\renewcommand{\arraystretch}{1.25}

\begin{tabular}{
  L{2.0cm}  
  L{1.8cm}  
  L{2.5cm}  
  L{1.0cm}  
  L{1.4cm}  
  L{1.6cm}  
  L{1.7cm}  
  L{2.0cm}  
  L{3.5cm}  
  L{1.0cm}  
  L{1.2cm}  
  L{1.9cm}  
}

\toprule
\makecell{\textbf{Method}\\\textbf{Name}} &
\makecell{\textbf{Contrast}} &
\makecell{\textbf{Optics}\\\textbf{Model}} &
\makecell{\textbf{Lang.}} &
\makecell{\textbf{Autodiff.}\\\textbf{Library}} &
\makecell{\textbf{Object}\\\textbf{Rep.}} &
\makecell{\textbf{Recon.}\\\textbf{Method}} &
\makecell{\textbf{Learning}\\\textbf{Architecture}} &
\makecell{\textbf{Imaging}\\\textbf{Types}} &
\makecell{\textbf{Blind}} &
\makecell{\textbf{Shift}\\\textbf{Variant}} &
\makecell{\textbf{Ref}} \\
\midrule

\makecell{\href{https://github.com/mehta-lab/waveorder}{WaveOrder}} &
\makecell{Fl, VFl, Ph,\\ Bir, Abs} &
\makecell{Part.\ coh.\ vect.,\\ single scattering} &
\makecell{Python} &
\makecell{PyTorch} &
\makecell{Voxel} &
\makecell{Tk} &
\makecell{PI-FC} &
\makecell{WF, LS, DPC, OPM,\\ Conf, Pol.\ oblique ill.} &
\makecell{Yes} &
\makecell{Yes} &
\makecell{This work} \\

\rowcolor{rowgray}
\makecell{\href{https://github.com/chromatix-team/chromatix}{Chromatix}} &
\makecell{General} &
\makecell{Scalar \& vect.,\\ multislice} &
\makecell{Python} &
\makecell{JAX} &
\makecell{Voxel,\\ INR} &
\makecell{General} &
\makecell{General} &
\makecell{WF, holo., ptych.,\\ miniscope, Conf} &
\makecell{Yes} &
\makecell{Yes} &
\makecell{\cite{deb_chromatix_2025}} \\

\makecell{\href{https://github.com/cell-observatory/aovift}{AOViFT}} &
\makecell{Fl} &
\makecell{Scalar} &
\makecell{Python} &
\makecell{Tensorflow} &
\makecell{Voxel} &
\makecell{RL} &
\makecell{ViT} &
\makecell{LLS, WF} &
\makecell{Yes} &
\makecell{Yes} &
\makecell{\cite{alshaabi_fourier-based_2025}} \\

\rowcolor{rowgray}
\makecell{\href{https://github.com/iksungk/CoCoA}{CoCoA}} &
\makecell{Fl} &
\makecell{Scalar} &
\makecell{Python} &
\makecell{PyTorch} &
\makecell{INR} &
\makecell{RL} &
\makecell{MLP} &
\makecell{WF, 2-photon} &
\makecell{Yes} &
\makecell{Yes} &
\makecell{\cite{kang_coordinate-based_2024}} \\

\makecell{\href{https://github.com/elgw/deconwolf}{Deconwolf}} &
\makecell{Fl} &
\makecell{Scalar} &
\makecell{C} &
\makecell{None} &
\makecell{Voxel} &
\makecell{RL+} &
\makecell{None} &
\makecell{WF} &
\makecell{No} &
\makecell{No} &
\makecell{\cite{wernersson_deconwolf_2024}} \\

\rowcolor{rowgray}
\makecell{\href{https://github.com/MeatyPlus/Richardson-Lucy-Net}{RLN}} &
\makecell{Fl} &
\makecell{Scalar} &
\makecell{Python} &
\makecell{Tensorflow} &
\makecell{Voxel} &
\makecell{RLN} &
\makecell{PI-CNN} &
\makecell{WF, LS, SIM,\\ diSPIM, Conf,\\ STED, iSIM} &
\makecell{No} &
\makecell{Yes} &
\makecell{\cite{li_incorporating_2022}} \\

\makecell{\href{https://github.com/apsk14/rdmpy}{RDM}} &
\makecell{Fl} &
\makecell{Scalar} &
\makecell{Python} &
\makecell{PyTorch} &
\makecell{Voxel} &
\makecell{LS} &
\makecell{U-Net} &
\makecell{WF, LS, miniscope,\\ endoscope} &
\makecell{No} &
\makecell{Yes} &
\makecell{\cite{kohli_ring_2025}} \\

\rowcolor{rowgray}
\makecell{\href{https://github.com/Biomedical-Imaging-Group/DeconvolutionLab2}{Decon...Lab2}} &
\makecell{Fl} &
\makecell{Scalar} &
\makecell{Java} &
\makecell{None} &
\makecell{Voxel} &
\makecell{Tk, RL,\\ FISTA, +} &
\makecell{None} &
\makecell{WF, Conf} &
\makecell{No} &
\makecell{No} &
\makecell{\cite{sage_deconvolutionlab2_2017}} \\

\makecell{\href{https://github.com/mehta-lab/waveorder}{PTI}} &
\makecell{Ph, Bir} &
\makecell{Part.\ coh.\ vect.} &
\makecell{Python} &
\makecell{None} &
\makecell{Voxel} &
\makecell{Tk} &
\makecell{None} &
\makecell{Pol.\ oblique ill.} &
\makecell{No} &
\makecell{No} &
\makecell{\cite{yeh_permittivity_2024}} \\

\rowcolor{rowgray}
\makecell{\href{https://github.com/tlambert03/microsim}{microsim}} &
\makecell{Fl} &
\makecell{Scalar} &
\makecell{Python} &
\makecell{None} &
\makecell{Voxel} &
\makecell{None} &
\makecell{None} &
\makecell{WF, Conf, SIM} &
\makecell{N/A} &
\makecell{N/A} &
\makecell{N/A} \\

\makecell{\href{https://github.com/deepinv/deepinv}{DeepInverse}} &
\makecell{General} &
\makecell{General} &
\makecell{Python} &
\makecell{PyTorch} &
\makecell{General} &
\makecell{General} &
\makecell{General} &
\makecell{MRI, CT, optics,\\ general} &
\makecell{Yes} &
\makecell{Yes} &
\makecell{\cite{tachella_deepinverse_2025}} \\

\bottomrule
\end{tabular}

\caption*{\textbf{Extended Data Table 1: Comparison of computational imaging and reconstruction frameworks.}
\textbf{Abbreviations:}
Fl = fluorescence;
VFl = vector fluorescence;
Ph = phase;
Bir = birefringence;
Abs = absorption;
Part.\ coh.\ vect. = partially coherent vectorial;
Tk = Tikhonov;
RL = Richardson–Lucy;
RL+ = accelerated RL;
RLN = Richardson–Lucy network;
LS = least squares;
FC = fully connected;
PI = physics-informed;
PI-CNN = physics-informed CNN;
INR = implicit neural representation;
MLP = multi-layer perceptron;
ViT = vision transformer;
WF = widefield;
LS (imaging) = light sheet;
DPC = differential phase contrast;
OPM = oblique-plane microscopy;
Conf = confocal;
Pol.\ obl.\ ill. = polarization-resolved oblique illumination;
LLS = lattice light sheet;
SIM = structured illumination microscopy;
iSIM = instant SIM;
diSPIM = dual-view inverted selective plane microscopy;
STED = stimulated emission depletion;
holo. = holography;
ptych. = ptychography.
}
\refstepcounter{exttab}
\label{exttab:method_comparison}
\end{sidewaystable*}

\begin{extendedfigure*}
  \centering
  \includegraphics[width=\textwidth]{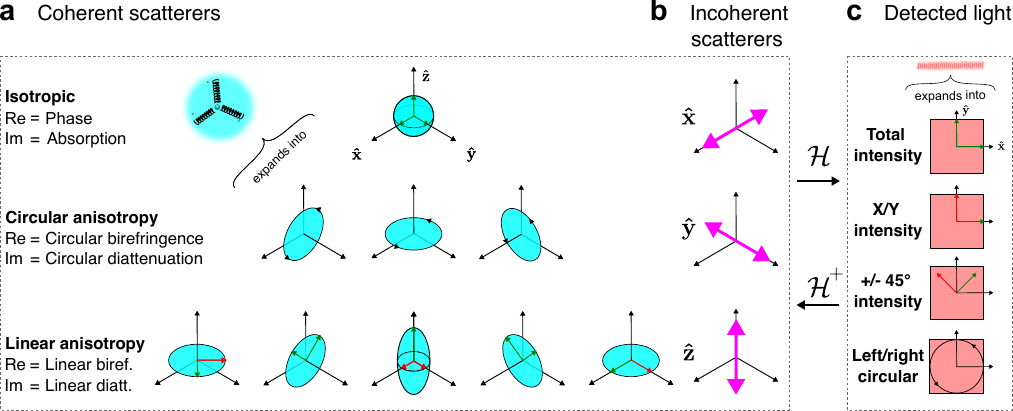}
  \caption{\textbf{Physically interpretable bases for modeling material properties and detected light. (a)} Coherent scatterers are described by a complex-valued scattering potential tensor that encodes how an incident field induces dipole moments within the specimen. Expanding this tensor into its spherical-harmonic components separates isotropic, circularly anisotropic, and linearly anisotropic responses (rows). The real parts of the expansion coefficients correspond to phase and birefringence, and the imaginary parts correspond to absorption and diattenuation. \textbf{(b)} Incoherent (fluorescent) scatterers are represented by emission dipoles oriented along $\mathbf{\hat{x}}$, $\mathbf{\hat{y}}$, or $\mathbf{\hat{z}}$. \textbf{(c)} Detected light is represented by Stokes parameters, allowing birefringence and diattenuation to be modeled within the same operator framework. The imaging operator linking \textbf{(a--b)} material properties to \textbf{(c)} measurable intensities is denoted $\mathcal{H}$, and its pseudo-inverse is denoted $\mathcal{H}^+$.} 
  \label{extfig:basis-functions}
\end{extendedfigure*}

\begin{extendedfigure*}
    \centering
    \includegraphics[width=\textwidth]{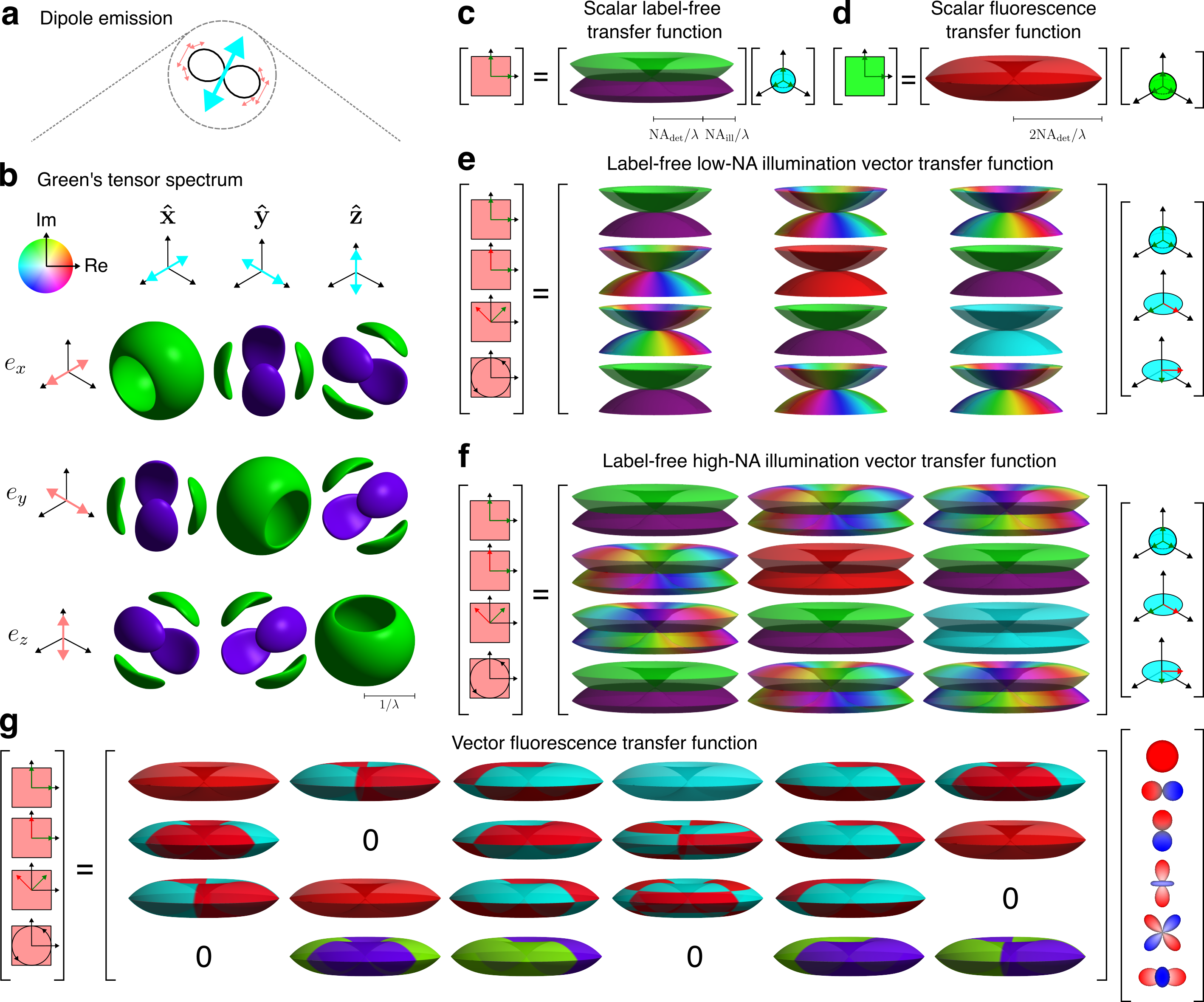}
    \caption{\textbf{Key components of the forward model.} \textbf{(a)} A dipole emitter (blue arrow) emits polarized fields (red arrows) in an anisotropic pattern (black line, radius proportional to power). \textbf{(b)} Arbitrary linear, circular, and elliptical dipole emissions can be modeled with the Green's tensor spectrum $G_{ij}(\nuvec)$. Each surface shows the quarter-maximum intensity of on-shell spectral field components (rows) $e_x$, $e_y$, and $e_z$ emitted by dipoles oriented along (columns) $\hat{\textbf{x}}$, $\hat{\textbf{y}}$, and $\hat{\textbf{z}}$, with relative phase shown in color (see color rose). Orthogonal slices through the Green's tensor spectrum and the Green's tensor are shown in \textbf{Supp.\,Fig.\,2c--g}. 3D support of various transfer functions with phase encoded in color. \textbf{(c)} Scalar label-free transfer function with $\text{NA}_{\text{ill}} = 0.5$ and $\text{NA}_{\text{det}} = 0.75$. \textbf{(d)} Scalar fluorescence transfer function with $\text{NA}_{\text{det}} = 0.75$. \textbf{(e--f)} Label-free vector transfer functions with $\text{NA}_{\text{det}} = 0.75$ and circularly polarized illumination that expresses polarization-resolved data (rows) as outputs of filters that modulate material properties (columns) with \textbf{(e)} low-NA illumination $\text{NA}_{\text{ill}} = 0.1$ and \textbf{(f)} high-NA illumination $\text{NA}_{\text{ill}} = 0.5$. \textbf{(g)} Vectorial fluorescence transfer functions with $\text{NA}_{\text{det}} = 0.75$ and circularly polarized illumination that expresses polarization-resolved data (rows) as outputs that modulate material properties (columns), here spherical harmonic components of fluorescent dipole orientation distribution functions.}
   \label{extfig:transfer}
\end{extendedfigure*}

\begin{extendedfigure*}
    \centering
    \includegraphics[width=\textwidth]{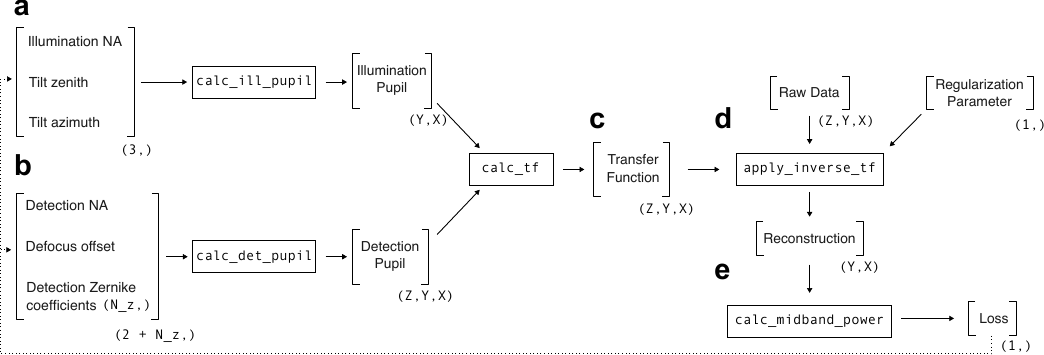}
    \caption{\textbf{Network architecture for auto-tuned phase-from-defocus reconstructions.}
\textbf{(a)} Illumination-parameter branch: illumination numerical aperture, tilt zenith, and tilt azimuth (3 parameters) are used to generate the illumination pupil ($\mathtt{Y,X}$) via \texttt{calc\_ill\_pupil}.
\textbf{(b)} Detection-parameter branch: detection numerical aperture, defocus offset, and detection Zernike coefficients (2 + \texttt{N\_z}) parameters are used to calculate the detection pupil (\texttt{Z,Y,X}) through \texttt{calc\_det\_pupil}.
\textbf{(c)} Transfer-function construction: illumination and detection pupils are combined using \texttt{calc\_tf} to form a 3D defocus transfer function (\texttt{Z,Y,X}).
\textbf{(d)} Inverse imaging model: the transfer function and raw defocused intensity stack (\texttt{Z,Y,X}) are passed to \texttt{apply\_inverse\_tf}, producing the phase reconstruction (\texttt{Y,X}) with a fixed regularization parameter.
\textbf{(e)} Mid-band power loss: reconstructed phase is evaluated by \texttt{calc\_midband\_power} to compute the scalar loss for optimization.
Arrows denote data flow; dashed arrows denote backpropagation to optimize the input parameters; dimensionalities are below each data object. This architecture jointly optimizes the phase reconstruction, illumination, and detection parameters from raw defocused measurements.}
    \label{extfig:optimization-arch}
\end{extendedfigure*}

\begin{extendedfigure*}
    \centering
\includegraphics[width=\textwidth]{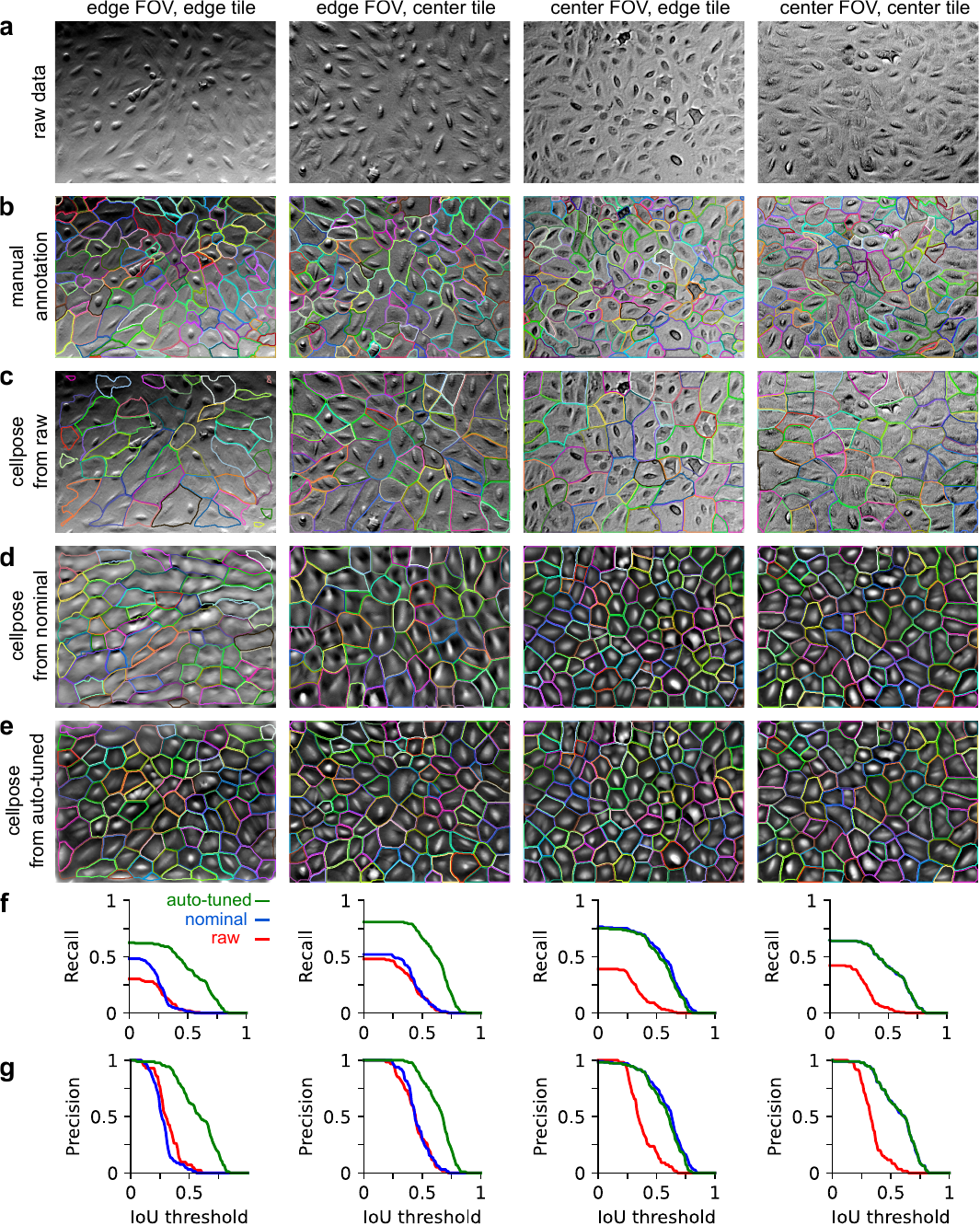}
    \caption{\textbf{Quantitative evaluation of auto-tuned reconstructions on optical pooled screen data. (a)} Raw label-free images from representative tiles across the 35~mm well show variation in illumination and contrast between center and edge tiles. \textbf{(b)} Manual segmentations used as ground truth. \textbf{(c)} CellPose segmentations on raw data. \textbf{(d)} Segmentations on nominal (untuned) reconstructions. \textbf{(e)} Segmentations on physics-informed, auto-tuned reconstructions. \textbf{(f--g)} Quantitative evaluation of segmentation accuracy, reported as recall and precision scores versus IoU threshold for raw (red), nominal (blue), and auto-tuned (green) reconstructions.}
    \label{extfig:ops-segment}
\end{extendedfigure*}

\begin{extendedfigure*}
    \centering
\includegraphics[width=125mm]{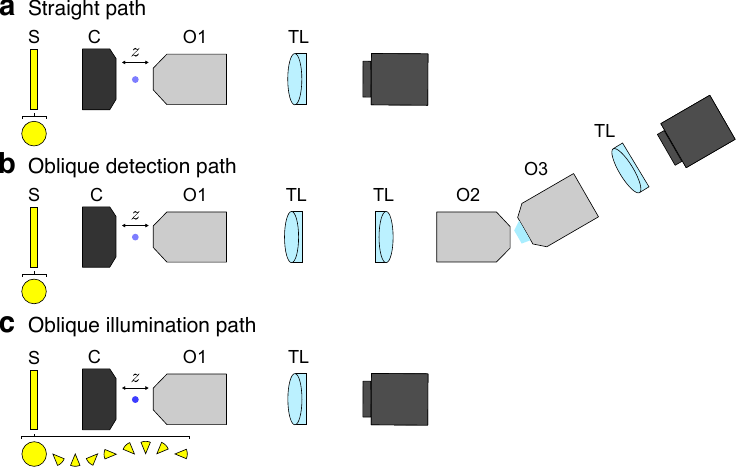}
    \caption{\textbf{Label-free light path schematics.}
\textbf{(a)} Straight path: light from the source (S) is focused by the condenser (C) and transmitted through the sample, which is moved along the $z$-axis. Transmitted light is collected by the first objective (O1) and imaged with a tube lens (TL) onto a camera.
\textbf{(b)} Oblique detection path: the illumination remains identical to \textbf{(a)}, but detection is re-imaged at an oblique angle using a secondary objective (O2) and tertiary objective (O3).
\textbf{(c)} Oblique illumination path: the detection path remains straight as in \textbf{(a)}, but illumination is provided obliquely by masked illumination sectors (yellow) in the source plane.
 S = source; C = condenser; O = objective; TL = tube lens. Blue dots indicate the sample plane; $z$ indicates the sample translation axis.}
    \label{extfig:light-paths}
\end{extendedfigure*}

\begin{extendedfigure*}
    \centering
\includegraphics[width=0.95\textwidth]{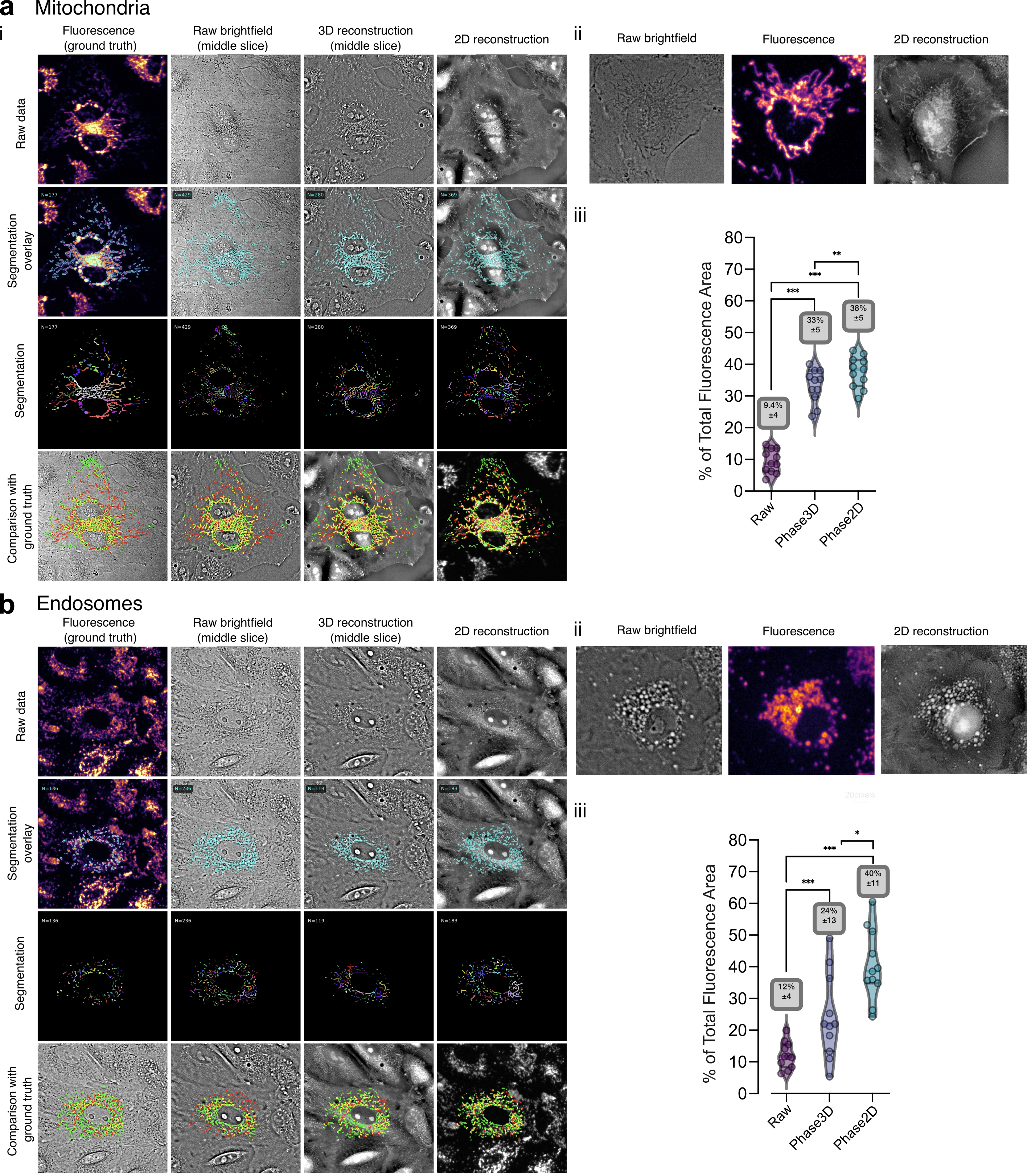}
    \caption{\textbf{Quantitative phase reconstruction enhances label-free organelle segmentation. (a)} Mitochondria segmentation in live A549 cells. \textbf{(i)} Comparison of imaging modalities: ground-truth fluorescence (mCherry–Mito), raw brightfield (center slice), 3D phase reconstruction (center slice), and 2D phase reconstruction (auto-tuned). Rows show raw intensity data, segmentation overlays (cyan), color-coded connectivity labels, and overlays of segmentation (yellow) between ground-truth fluorescence (red) and reconstructions (green). \textbf{(ii)} Zoomed regions highlight the recovery of fine mitochondrial networks in phase reconstructions relative to raw brightfield. \textbf{(iii)} Quantification of segmentation overlap, reported as the percentage of the total fluorescent area covered by the label-free segmentation ($N=11$ cells). 2D phase reconstruction (38\% $\pm$ 5) significantly outperforms raw brightfield (9.4\% $\pm$ 4). \textbf{(b)} Endosome segmentation in A549 cells. \textbf{(i)} Comparison of imaging modalities as in (a) using GFP-labeled endosomes. \textbf{(ii)} Zoomed regions show improved contrast of vesicular structures in raw and phase-reconstructed images. \textbf{(iii)} Quantification of segmentation overlap shows similar improvements for endosomes ($N=11$ cells). Statistical analysis used repeated-measures ANOVA with post-hoc paired $t$-tests and Bonferroni correction. Box plots indicate mean $\pm$ s.d.; significance levels: (*) $p<0.05$, (**) $p<0.01$, (***) $p<0.001$. \textbf{Scale bars:} 20~\textmu m.}
    \label{extfig:organelle-segmentation}
\end{extendedfigure*}

\begin{extendedfigure*}
    \centering
\includegraphics[width=\textwidth]{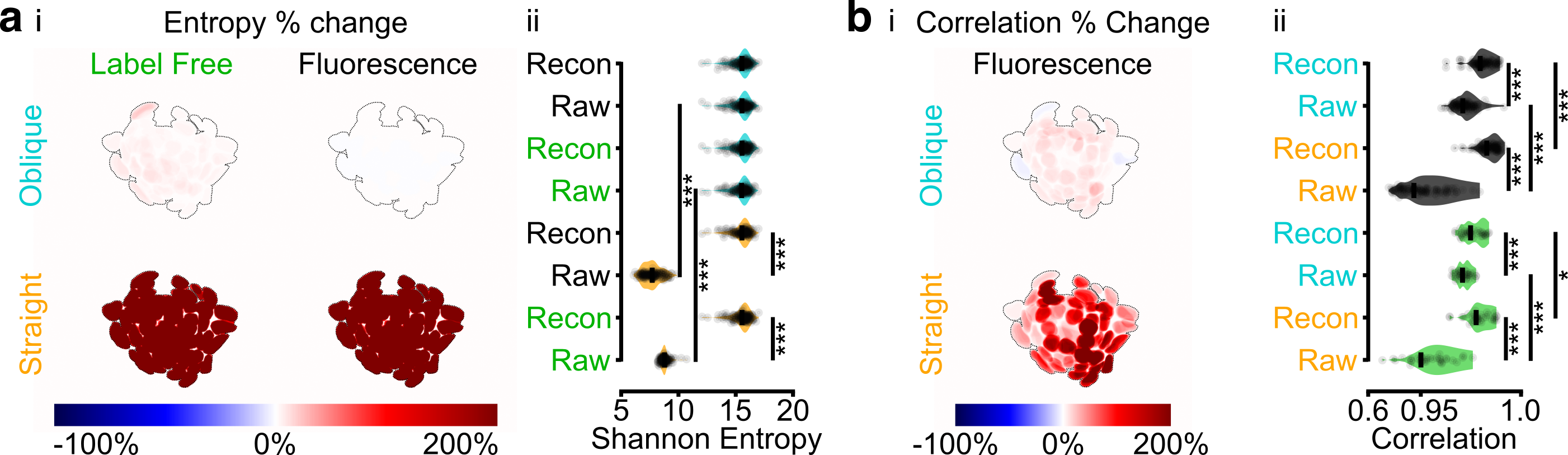}
    \caption{\textbf{Neuromast cell classification.} 
    \textbf{(a)(i)} Shannon Entropy \% change during reconstruction of oblique label-free detection (top left), oblique fluorescence detection (top right), straight label-free (bottom left) and straight fluorescence (bottom right). 
    \textbf{(a)(ii)} Violin plot showing the distributions of single cell Shannon Entropy values for straight and oblique detection both before and after reconstruction.
    \textbf{(b)(i)} Correlation \% change for oblique (top) and straight (bottom) detection during reconstruction.
    \textbf{(b)(ii)} Violin plot showing the distributions of single cell correlation values for mantle (green) and hair/support (black) cells imaged with straight or oblique detection both before and after reconstruction.}
    \label{extfig:neuromast-segmentation}
\end{extendedfigure*}

\begin{extendedfigure*}
    \centering
\includegraphics[width=\textwidth]{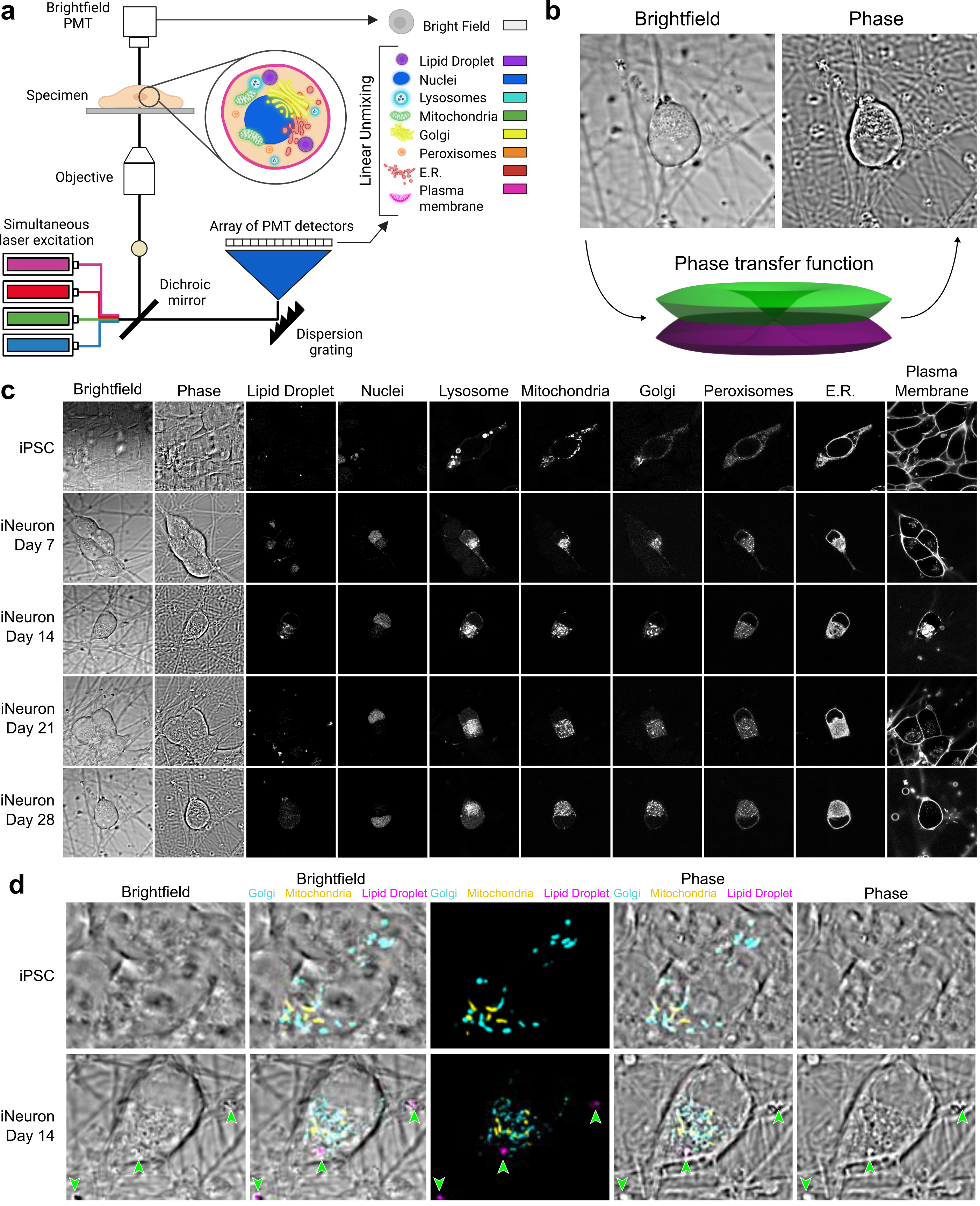}
    \caption{\textbf{Multispectral iPSC data. (a)} Configuration for multispectral imaging of iPSCs and iPSC-derived neurons (iNeurons), combining a label-free channel with multiple fluorescence channels. Linear unmixing enables organelle-specific interpretation. \textbf{(b)} Representative label-free images (left) and reconstructed phase. \textbf{(c)} Restored and unmixed images (columns) across differentiation stages (rows), demonstrate a rich mapping between the dense label-free channel and the unmixed fluorescence channels. \textbf{(d)} Comparing brightfield (far left) and reconstructed phase (far right) to three fluorescence organelle channels (center) with overlays (center left and right) show improved correlation between reconstructed phase and fluorescence, particularly in the lipid droplet channel (green arrows).}
    \label{extfig:multispectral-ipsc}
\end{extendedfigure*}

\begin{extendedfigure*}
    \centering
    \includegraphics[width=\textwidth]{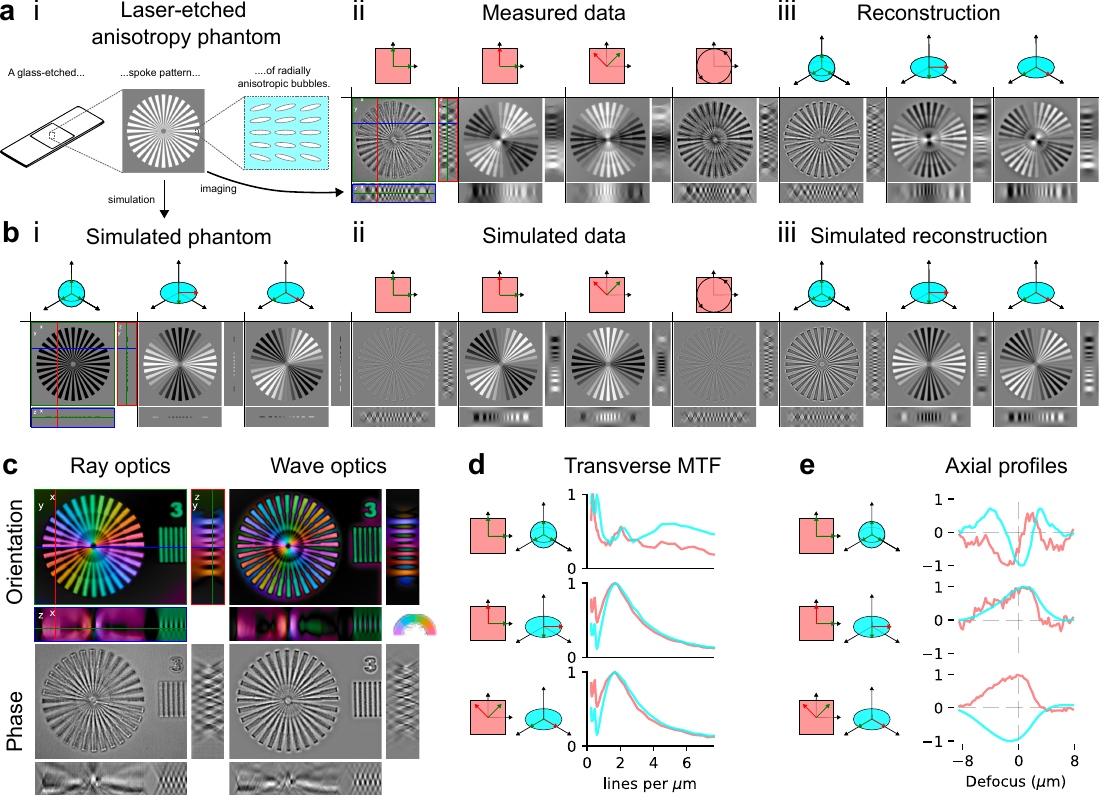}
    \caption{\textbf{Demonstration of vectorial multi-channel reconstruction with experiment and simulation.} We imaged \textbf{(a)} \textbf{(i)} a laser-etched spoke pattern of transverse radially anisotropic bubbles. \textbf{(ii)} We made volumetric measurements of the Stokes parameters (columns), and applied a multi-channel label-free reconstruction to \textbf{(iii)} recover material properties. \textbf{(b)} We \textbf{(i)} simulated the phantom's material properties, \textbf{(ii)} simulated the imaging process, then \textbf{(iii)} simulated the reconstruction. \textbf{(c)} We encoded transverse birefringence properties into color, where brightness indicates the strength of the anisotropy and hue indicate the slow-axis orientation. We compared the orientation and phase reconstructions (rows) in ray- and wave-optics reconstructions (columns). \textbf{(d)} For each data and material property, we measured a series of azimuthal profiles at different radii on the spoke pattern and used the 10th--90th percentile modulation as an empirical estimate of the transverse modulation transfer function (MTF). \textbf{(e)} Similarly, we measured axial profiles through each data and material property. See also \textbf{Video 5}.}
    \label{extfig:vector_quantitative_reconstruction}
\end{extendedfigure*}
\clearpage

\end{refsegment}

\defbibfilter{newinmethods}{
  segment=2 and not segment=1
}
\printbibliography[filter=newinmethods,title={References},resetnumbers=false]

\section{Biological materials availability}
All unique biological materials classified are available from authors upon reasonable request or from commercial sources.

\section{Data availability}
Demo data with example reconstructions are available in the \href{https://github.com/mehta-lab/waveorder}{WaveOrder} repository.

\section{Code availability}
Code can be found in the \href{https://github.com/mehta-lab/waveorder}{WaveOrder} repository. 

\section{Acknowledgements}
We thank Abi Koh and \href{https://github.com/shashmehta}{Shashvat Mehta} for contributing to \textbf{Video 1} and \textbf{Video 3}, respectively. We thank Taylla Milena Theodoro for supporting software pipelines. We thank Li-Hao Yeh for contributing the cardiomyocyte data. We thank the Biohub Scientific Computing Platform for enabling high-performance reconstructions. All authors are supported by intramural funding from Biohub. We thank Priscilla Chan and Mark Zuckerberg for supporting Biohub. S.B.M. and S.C. acknowledge support from Allen Distinguished Investigator Program  for the collaborative work on extending WaveOrder to multi-spectral imaging of iPSC differentiated cells.

\section{Author contributions}
\textbf{Conceptualization:} T.\,C., S.\,B.\,M.
\textbf{Methodology:} T.\,C., G.\,S., Z.\,L., A.\,H., A.\,Q.\,R., S.\,N.\,R., S.\,B.\,M.
\textbf{Sample preparation \& imaging:} T.\,C., I.\,E.\,I., E.\,H.-M., S.\,X., X.\,Z., D.\,S., A.\,Q.\,R., M.\,S., C.\,L., S.\,N.\,R., M.\,C.\,Z., S.\,P., S.-C.\,L.
\textbf{Software:} T.\,C., Z.\,L., I.\,T., A.\,V., S.\,R.\,V.
\textbf{Writing---draft:} T.\,C., S.\,B.\,M.
\textbf{Writing---review \& editing:} all authors.
\textbf{Supervision:} S.\,C., C.\,A., M.\,D.\,L., A.\,J., K.\,B., L.\,A.\,R., S.\,B.\,M.

\section{Competing interests}
The authors declare no conflicts of interest.

\end{document}